\documentclass[acmsmall,screen]{acmart}

\AtBeginDocument{%
  \providecommand\BibTeX{{%
    \normalfont B\kern-0.5em{\scshape i\kern-0.25em b}\kern-0.8em\TeX}}}

\setcopyright{acmcopyright}
\copyrightyear{2020}
\acmYear{2020}

\renewcommand\footnotetextcopyrightpermission[1]{} %

\usepackage{multirow}
\usepackage{booktabs}
\usepackage{ifthen}
\usepackage{color}
\usepackage{url}
\usepackage{xspace}
\usepackage[T1]{fontenc}
\usepackage{enumerate}
\usepackage[linesnumbered,ruled,vlined]{algorithm2e}
\usepackage{array,multirow,graphicx}
\usepackage{float}
\usepackage{balance}
\usepackage{tikz}
\usepackage{calc}
\usepackage{subfigure}
\usepackage{listings,amsfonts}
\usepackage{amsmath,xcolor,pifont}
\usepackage{url}
\usepackage{xspace}
\usepackage{hyperref,endnotes}
\usepackage{xcolor}
\usepackage{array}
\usepackage{multirow,makecell}
\usepackage[misc]{ifsym}
\usepackage{enumitem}

\definecolor{yellow}{RGB}{255,255,153}
\definecolor{grey}{RGB}{224,224,224}

\begin{document}

\title{A Tale of Two Countries: A Longitudinal Cross-Country Study of Mobile Users' Reactions to the COVID-19 Pandemic Through the Lens of App Popularity}
\renewcommand{\shorttitle}{A Tale of Two Countries}

\author{Liu Wang}
\affiliation{%
  \institution{Beijing University of Posts and Telecommunications}
  \city{Beijing}
  \country{China}
}

\author{Haoyu Wang}
\authornote{Corresponding Author: Haoyu Wang (haoyuwang@bupt.edu.cn).}
\affiliation{%
  \institution{Beijing University of Posts and Telecommunications}
  \city{Beijing}
  \country{China}
}
\email{haoyuwang@bupt.edu.cn}

\author{Yi Wang}
\affiliation{%
  \institution{Beijing University of Posts and Telecommunications}
  \city{Beijing}
  \country{China}
}

\author{Gareth Tyson}
\affiliation{%
  \institution{Queen Mary University of London}
  \city{London}
  \country{United Kingdom}
}

\author{Fei Lyu}
\affiliation{%
  \institution{Beijing University of Posts and Telecommunications}
  \city{Beijing}
  \country{China}
}

\begin{abstract}
The ongoing COVID-19 pandemic has profoundly impacted people's life around the world, including how they interact with mobile technologies. In this paper, we seek to develop an understanding of how the dynamic trajectory of a pandemic shapes mobile phone users' experiences. Through the lens of app popularity, we approach this goal from a cross-country perspective. We compile a dataset consisting of six-month daily snapshots of the most popular apps in the iOS App Store in China and the US, where the pandemic has exhibited distinct trajectories. Using this longitudinal dataset, our analysis provides detailed patterns of app ranking during the pandemic at both category and individual app levels. We reveal that app categories' rankings are correlated with the pandemic, contingent upon country-specific development trajectories. Our work offers rich insights into how the COVID-19, a typical global public health crisis, has influence people's day-to-day interaction with the Internet and mobile technologies.   

\end{abstract}

\maketitle

\section{Introduction}

The ongoing COVID-19 pandemic has impacted almost every aspect of human life. The pandemic introduced changes in people's perceptions and attitudes, as well as how they work, pursue leisure, interact with others, and so on~\cite{bussing2020tumor,dwivedi2020impact,kramer2020potential}. Among these changes, interacting with internet and information technologies could be very significant. Compared with prior years, technologies have played a critical role in the pandemic, as an essential infrastructure for supporting the response to the pandemic. For instance, to help curb its spread, governments and public health authorities around the world have launched contact-tracing apps~\cite{cho2020contact,ahmed2020survey}. Given that mobile phones have been the most popular and pervasive interfaces for people to interact with the Internet, most technological experience has occurred via mobile apps. We therefore posit that \textit{the popularity of mobile apps, as well as their dynamics, may provide a lens for understanding people's experiences with technologies during the pandemic.} 

Due to the different strategies employed by countries in fighting COVID-19, as well as the degree of public compliance, the regional development of the pandemic has exhibited distinct patterns. For example, China introduced a strict nation-wide lockdown during the initial outbreak and achieved strong public cooperation, making the pandemic largely under control in most of areas from April~\cite{altakarli2020china}. In contrast, the United States, although implementing multiple measures at different levels, failed to contain the pandemic, resulting in chaos spanning the entire year~\cite{CasesinUS}. We conjecture that such differences may be reflected in the app popularity in different countries. \textit{Thus, understanding people's experiences with mobile apps must consider the country-specific patterns of the pandemic.}

\textbf{This Work.}
In this paper, we seek to understand, characterize, and compare the dynamics of app popularity during the pandemic, under the circumstance of different country-specific pandemic development patterns.
We are particularly keen to explore the relationship between app utilization and the infection trajectories experienced across countries. 
We first create the daily snapshots of the ranking of popular apps in the iOS App Store in both China and the United States (US), from the January 2020 to the end of June 2020 (\textbf{See Section~\ref{sec:design}}). 
This longitudinal dataset enables us to study app usage behavior evolution across billions of mobile users in both China and US.
We next characterize the dynamics of app popularity from both a Macro- (i.e., app category) and Micro- (i.e., individual app) lens. 
We investigate which app categories are positively or negatively correlated with the coronavirus pandemic and try to formulate general laws in their ranking evolution (\textbf{See Section~\ref{sec:category}}), especially when comparing the differing situations and cultures in China and the US. 
We further perform fine-grained per-app analysis (\textbf{See Section~\ref{sec:perapp}}), to summarize the dynamics of app popularity.
Finally, we investigate whether the COVID-19 pandemic has introduced side-effects on the app maintenance behaviors (e.g., performance) by analyzing millions of app reviews (\textbf{See Section~\ref{sec:sideeffect}}).
We highlight a number of key findings:

\begin{itemize}
    \item \textit{COVID-19 has played a key role in changing the popularity of some categories of app}. During the outbreak, several app categories experienced significant ups and downs in popularity in both China and the US, particularly \texttt{Business}, \texttt{Education}, \texttt{Navigation}, and \texttt{Travel}. This change reflects the reactions and efforts people made in response to the outbreak, and highlights differences between China and the US.
    
    \item \textit{The impact of COVID-19 on app rankings are diverse and can vary from country to country}. There are many apps in our dataset whose rankings are strongly correlated, either positively or negatively, with the pandemic case rates. Their ranking variations can be classified into several groups. Most apps within the same group share some similarities in their adaptation to the pandemic situation.
    
    \item \textit{The rapid popularity of some apps may be accompanied by the side effect of declining ratings, especially for Chinese apps}. We explore the reasons for this from the app reviews and found a massive increase in the number of negative reviews compared to the pre-popularity. We reveal some potential challenges for apps.
\end{itemize}

To the best of our knowledge, this is the first longitudinal study to characterize the behavior dynamics of mobile users through the lens of mobile app popularity. We have released our dataset and the experimental results to the research community at Github:

\begin{center}
    \url{https://app-popularity-covid19.github.io/}
\end{center}

\section{Background and Related Work}
\label{sec:background}

\subsection{COVID-19 Pandemic in China and the United States}

We start by describing the development of COVID-19 pandemic and contrasting the responses of China and the US. 

\subsubsection{COVID-19 in China}

The first case of COVID-19 was identified in Wuhan, China, in December 2019.
In the face of a previously unknown virus, China rolled out perhaps the most ambitious, agile, and aggressive disease containment effort in history~\cite{whoreport}. The central and local governments launched the national emergency response and various prevention and control measures have been implemented rapidly. Investigations began quickly and traced the outbreak to a seafood market, which was immediately closed by Chinese authorities as an initial method to terminate all meat trades. 

In response to the rapid spread of SARS-Cov-2 within Hubei province, the Chinese government expanded its precautionary measures, announcing a complete lockdown of Wuhan and Hubei province cities by closing airports, railway stations, highway entrances, and suspending all local public transportation to prevent anyone from entering or leaving~\cite{10.1093/jtm/taaa036}. The government also implemented aggressive quarantine and social distancing in the entire country including cancelling activities with large crowds, wearing a mask when going out, postponing the reopening of schools, enabling home office, etc. After all the measures taken and people’s full commitment, a decline in the number of new cases and deaths was clearly observed by the end of February. Later on,  with all the situation improvements happening, China announced the lifting of the embargo and travel restrictions on Wuhan and restarting of the economy in April~\cite{CNBC}.

\subsubsection{COVID-19 in US}

The first case of COVID-19 in the US was reported on January 20, 2020~\cite{doi:10.1056/NEJMoa2001191} and the first known American deaths occurred in February~\cite{CNN}. President Donald Trump declared the US outbreak a public health emergency on January 31. Subsequently, restrictions were placed on flights arriving from China~\cite{public-health-emergency,travel-claim}. However, the initial US response to the outbreak was otherwise slow in terms of preparing the medical system, halting other travel, and testing~\cite{apnews,politico,theguardian}. Meanwhile, Trump downplayed the threat posed by the virus and claimed the outbreak was under control~\cite{trump-timeline}. In addition, each state that had imposed a stay-at-home order or shelter in place had begun lifting the restrictions of businesses and public spaces in May~\cite{reopening}. 
Despite its considerable advantages --- abundant resources, biomedical infrastructure, and scientific expertise --- the US has been severely hit by COVID-19. More than 21 million confirmed cases have been reported by the time of writing, resulting in more than 368K deaths, which is the most of any country and the 14th highest on a per capita basis~\cite{jhu-csse,jhu}.

\iffalse
Figure~\ref{fig:trends} shows the trends in the number of new confirmed cases per day in the two countries, from which we can see a shape contrast.

\begin{figure}[htbp]
    \centering
    \subfigure[China]{
    \includegraphics[width=0.48\textwidth]{Figures/cn-trend.pdf}}
    \subfigure[US]{
    \includegraphics[width=0.48\textwidth]{Figures/us-trend.pdf}}
    \vspace{-0.15in}
    \caption{The trends in the number of daily new cases in China and the US.}
    \label{fig:trends}
\end{figure}
\fi

\subsection{App Ranking in iOS App Store}
App ranking refers to the position of an app in a store or marketplace. In general, app rankings can reference not only how an app ranks in a search, but also the overall app rankings from the store. These rankings are country specific, and can be viewed at an overall or category level. The Apple App Store, one of the most common app marketplaces, provides the real-time app ranking lists both overall and by category. In the world of apps, ranking is of utmost importance. A minor change in rank can make a significant difference in traffic and revenue. Apptentive's recent mobile consumer survey~\cite{survey} reveals that nearly half of all mobile app users identified browsing the app store charts and search results (the placement on either of which depends on rankings) as a preferred method for finding new apps. Simply put, higher rankings mean more downloads, and to some degree, more money for app developers.

The Apple iOS App Store has a complex and highly guarded algorithm for determining rankings for both keyword-based app store searches and composite top charts. Although the exact ranking algorithms are not publicly available, there are several known factors that influence the rankings, including downloads (how many people have downloaded the app), rating (how many stars people give the app), rating count (how many ratings the app has) and trends (how quickly the app is growing), etc~\cite{factors}.
\textit{Since it is the combination of several other factors (average rating, number of ratings, installs, etc.), this ranking speaks to the overall awareness, usefulness, and satisfaction of an app. Consequently, app ranking is a particularly useful metric for mobile app analysis, particularly, its popularity.}

\subsection{Related Studies on COVID-19}
Since its outbreak, COVID-19 has attracted much attention from various research communities. A large number of studies are focused on the medical domain. Many medical researchers have made tremendous contributions to  pathology study, epidemiology study, treatment study and so on~\cite{bai2020presumed,zu2020coronavirus,onder2020case,shen2020treatment}.
Also, a number of computer scientists have adopted computing techniques like machine learning to help medical practitioners cope with the disease~\cite{latif2020leveraging}. 
For example, Zhang et al.~\cite{zhang2020covid} proposed the confidence-aware anomaly detection (CAAD) model to screen viral pneumonia on chest X-ray images. Wang et al.~\cite{wang2020covid} proposed COVID-Net, a deep convolutional neural network design tailored for the detection of COVID-19 cases from chest X-ray (CXR) images.

There are a growing number of mobile app studies relevant to COVID-19. Ahmed et al.~\cite{9144194} provided the first comprehensive review of contact tracing apps and discussed the concerns users have reported regarding their usage.
Oliver et al.~\cite{oliver2020mobile} described how mobile phone data can guide government and public health authorities in determining the best course of action to control the COVID-19 pandemic and in assessing the effectiveness of control measures.
He et al.~\cite{he2020beyond} presented a systematic analysis of coronavirus-themed mobile malware and found these apps aim to steal users’ private information or to make a profit by phishing and extortion.
There is also research into security and privacy issues. For example, Sun et al.~\cite{sun2021empirical} proposed an automated assessment tool to determine security and privacy weaknesses for apps, and undertook a user study to investigate concerns regarding contact tracing apps. Sharma et al.~\cite{sharma2020advocating} assessed privacy controls offered in COVID-19 apps and users' preferences if they were to adopt a COVID-19 app. Wang et al.~\cite{wang2020market} reveals various issues related to contact tracing apps from the users' perspective by analyzing a large number of user reviews.
Although a few previous studies have looked at COVID-19 through the lens of mobile apps, to the best of our knowledge, our work is the first comprehensive study to investigate mobile users' reactions to the COVID-19 pandemic through the lens of app ranking.

\subsection{Related Work on the Mobile App Ecosystem}
A large number of research studies have characterized the mobile app ecosystem from different perspectives. Wang et al.~\cite{wang2019understanding} characterized the evolution of Google Play based on 5.3 million app records collected from three snapshots of Google Play over three years. They observed that although the overall ecosystem shows promising
progress, there still exists a considerable
number of unsolved security issues. Besides, some studies were focused on app developers~\cite{wang2017explorative,wang2019characterizing,wang20203}, app market maintenance behaviors~\cite{RMVDroid,wang2018android,wang2018beyond}, app clone detection~\cite{li2019rebooting,li2019identifying,wang2015using}, third-party libraries~\cite{zhan2020automated,libradar,dong2018mobile,wang2017understanding,xu2020dissecting}, security~\cite{zhou2012dissecting,maddroid,liu2019dapanda,dong2018frauddroid,hu2020mobile,zhou2020demystifying,tang2020all,wang2019signing,hu2018dating} and privacy issues~\cite{razaghpanah2018apps,reardon201950,xi2019deepintent,wang2017understanding,liu2016identifying,wang2015reevaluating} of the ecosystem, based on app binary analysis, UI analysis, app traffic analysis, and app metadata analysis, etc.
This paper, investigates the mobile app ecosystem from a novel perspective, to understanding how the dynamic trajectory of the COVID-19 pandemic shapes mobile smartphone users' experiences.

\section{Study Design}
\label{sec:design}

\subsection{Research Questions}
Our study is driven by the following research questions:

\begin{itemize}
\item[RQ1] \textbf{General Trends.}
\emph{Across different regions, what are the characteristics of app popularity during COVID-19, and how are they impacted by regional measures?}
%     We seek to study the popularity (i.e., the number of top-ranked apps) of different app categories, understand which app categories are more competitive over the outbreak and explore their general laws in ranking evolution.
    
%     \item[RQ2] 
%     \textit{What is the correlation between app ranking trends and the evolution of the pandemic situation?}
%     Considering that the outbreak may affect people's usage of mobile apps, it is thus interesting to investigate what kinds of apps are positively or negatively correlated with the coronavirus pandemic.
    
\item[RQ2] \textbf{Behavior Patterns.}
\textit{What are the different patterns in the change of app popularity over the course of the pandemic?} 
We attempt to classify the app ranking variations according to relevance to the pandemic, then examine the distribution of the number of apps in each class and explore their characteristics.

\item[RQ3] \textbf{Side-effect.}
%The COVID-19 pandemic has drastically changed working and social habits for billions of people, which could be reflected on the mobile app usage.
\textit{Considering that app popularity may change greatly during the COVID-19 pandemic, does it introduce any side-effect on app maintenance behaviors?} 

\end{itemize}

\subsection{Data Collection}

We first harvest a dataset of app popularity. Given that the most straightforward impact of COVID-19 on the mobile app ecosystem will be mirrored in a number of popular apps, we leverage the App Store's ranking list and focus on the top-ranked apps. However, it is non-trivial to obtain such a dataset as it requires effort to continuously monitor apps in the app market to check their rankings. To this end, we implement a crawler to retrieve the daily app ranking list from iOS app store. Considering that the pandemic in different countries can evolve in different ways, we take two representative countries (China and the US), as case studies for comparison. %to make our characterization more robust.

To be specific, we crawl the ranking list with top 1500 apps each day from January 1 to June 30, 2020, to keep track of the apps and their daily rankings, containing over 543K records in total for the two countries. We then take the top 100 apps per day as the target of this study. 
Overall, there are 586 unique apps in China and 590 apps in the US covering 22 categories, each of which ranked in the top 100 for at least one day during a given half year.
Besides, we collect ratings for each app on a daily basis, as well as the reviews for some of these apps. 

Table~\ref{tab:dataset} summarizes our dataset.

\begin{table}[htbp]
    \centering
    \caption{Overview of our dataset (from Jan 1 to June 30, 2020).}
    \begin{tabular}{c|ccccc}
    \toprule
     Country & \# Categories & \# Top100 Unique Apps & \# Ranking Records & \# Ratings & \# Reviews \\
     \hline
     China & 22 & 587 & 271,765 & 106,653 & 980,573 \\
     US & 22 & 590 & 271,522 & 107,381 & 756,617 \\
     \bottomrule
    \end{tabular}
    \label{tab:dataset}
\end{table}

\subsection{Analysis Preliminaries}

We briefly present the methodologies used in the subsequent sections.

\subsubsection{Popularity of app categories.}
To understand which app categories are more competitive and explore general laws in ranking evolution, we begin our analysis by investigating the most intuitive metric of app-category popularity, i.e., the number of apps ranked in the top 100 for each category. 
Specifically, we measure the popularity of category by app volume, i.e., the number of apps belonging to that app category in the top-100. The larger the number, the more popular the category is. We examine the number of apps in each category on a daily basis to gain insight into how their popularity trends are changing. In more detail, we also zoom in on each category to check the ranking variations of all apps in that category.

\subsubsection{Correlation analysis.}
To understand if any apps have been significantly affected by the pandemic, and whether the impact has been positive or negative, we also study the correlation between the pandemic evolution and the app ranking changes, which could reflect the popularity changes of apps, for each app.
In our case, we measure the development of the pandemic situation over the half year through two metrics, i.e., the number of daily new confirmed cases and daily active cases, respectively. Moreover, we present a correlation comparison strategy based on Pearson Correlation Coefficient~\cite{pearson}. The Pearson Correlation Coefficient is a number that summarizes the direction and closeness of linear relations between two variables, taking a value between -1 and 1. The closer the correlation coefficient is to 1 (-1), the stronger the positive (negative) correlation is.

For each app, the daily ranking is viewed as a vector in Pearson Correlation Coefficient formula, then the correlation degree is obtained separately from the daily number of newly diagnosed cases and active cases in the corresponding countries. 
Note that we also take P-value into account because it tells us whether the result of an experiment is statistically significant. Typically if the P-value is lower than the conventional 5\% (P<0.05), the correlation coefficient is called statistically significant. Thus, results with P-value less than 0.05 are of concern to us, otherwise are ignored as insignificantly correlated.

Finally, we calculate the correlation coefficient for each app and obtained 358(61.1\%) apps in China and 497(84.2\%) apps in the US with statistically significant correlation with the number of daily new cases in respective countries, as well as 425(72.5\%) apps in China and 502(85.1\%) apps in the US with statistically significant correlation with daily active cases.
The results illustrate that the majority of apps have a statistically significant correlation between their rankings and the COVID-19 case rates.
This encourages and motivates the following analysis, i.e., the distributions of these correlation coefficients at the overall and category level. Due to the greater number of apps with statistically significant correlation between rankings and active cases compared to newly confirmed cases, we take the active cases as the pandemic marker for the subsequent study. And note that all the correlation coefficients provided below are statistically significant.

\iffalse
\begin{table}[htb]
    \centering
    \caption{The number of apps whose ranking has a significant correlation coefficient with the daily new cases or daily active cases}
    \vspace{-0.1in}
    \resizebox{0.7\textwidth}{!}{
    \begin{tabular}{l|c|c}
    \toprule
         & \# with Daily New cases & \# with Daily Active cases \\
         \midrule
    China & 358 (61.1\%) & 425 (72.5\%) \\
    \hline
    US & 497 (84.2\%) & 502 (85.1\%) \\
    \bottomrule
    \end{tabular}}
    \label{tab:number}
\end{table}
\fi

\section{Popularity of App Categories During the Pandemic}
\label{sec:category}

\begin{figure*}[h]
    \centering
    \subfigcapskip=-5pt
    \begin{center}
        \subfigure{
        \includegraphics[width=1\textwidth]{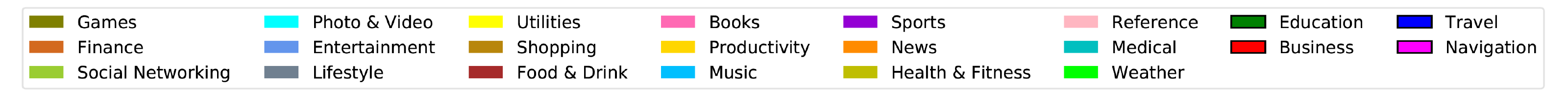}
        } 
        \vskip -0.05in
        \setcounter{subfigure}{0}
        \subfigure[China]{
        \includegraphics[width=1\textwidth]{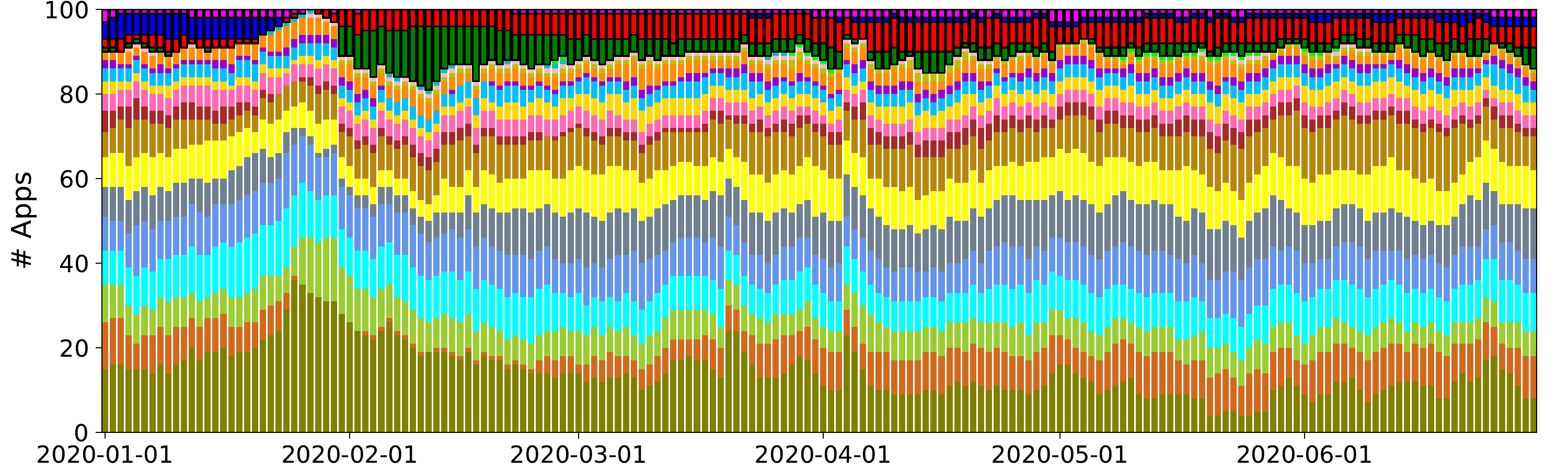}
        \label{fig:category-cn}}
        \vskip -0.03in
        \subfigure[US]{
        \includegraphics[width=1\textwidth]{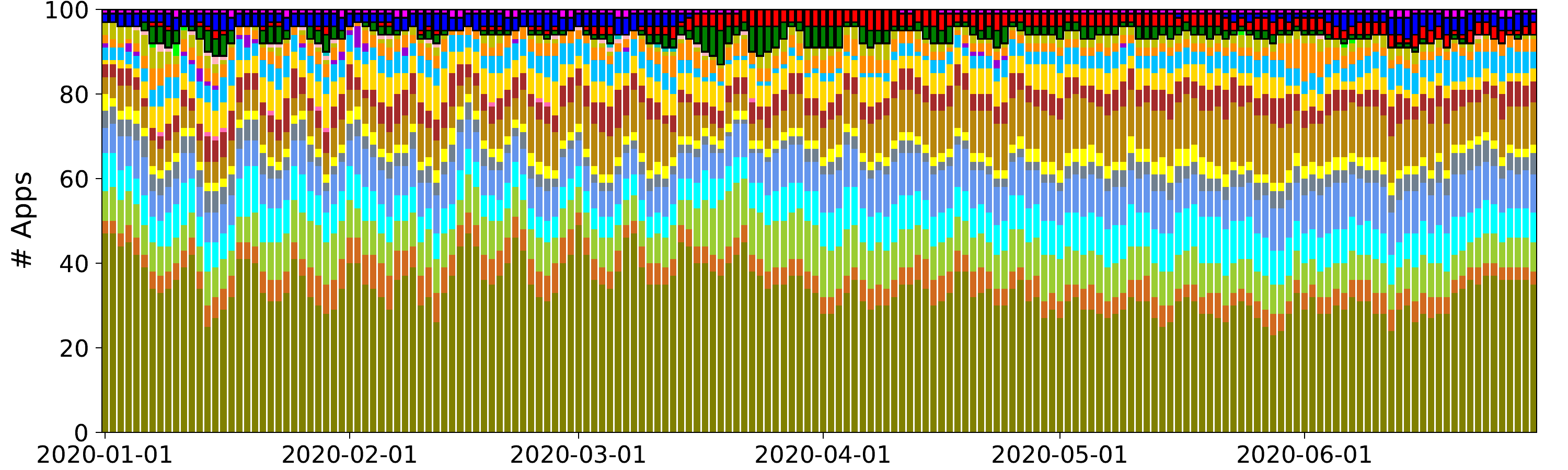}
        \label{fig:category-us}}
        \vspace{-0.15in}
        \caption{The distribution of number of apps in each category among top-100 apps per day}
        \label{fig:category}
    \end{center}
\end{figure*}

\subsection{The dynamics of the popularity}
\subsubsection{Coarse-grained inspection.}
Figure~\ref{fig:category} visualizes the dynamics of the popularity of each app category. We use stacked bar charts to show the number of apps in each category among each day's top 100 apps in the app stores of China and the Unite States, respectively. As aforementioned, there are 22 categories (see the legend in Figure~\ref{fig:category}). 
In a specific day, the change in  the numbers of apps in each category over time is represented by the fluctuations of the vertical lengths of the corresponding color.

In general, most category's popularity does not exhibit significant fluctuations in either China or the US. Some categories has been constantly popular over the long run, such as \texttt{Games}, \texttt{Social Networking}, \texttt{Photo \& Video} and \texttt{Entertainment}, as well as some being consistently less popular, such as \texttt{Reference}, \texttt{Medical} and \texttt{Weather}. However, there are four categories that are worth noting in both countries, i.e., \texttt{Business}, \texttt{Education}, \texttt{Travel}, and \texttt{Navigation}, whose popularity exhibits 
remarkable dynamics along with the outbreak in the respective countries. 
We further examine the details of the dynamics of their popularity.

\subsubsection{Fine-grained inspection.}
\begin{figure*}
    \centering
    \subfigure[Business]{
    \label{fig:business}
    \includegraphics[width=0.48\textwidth]{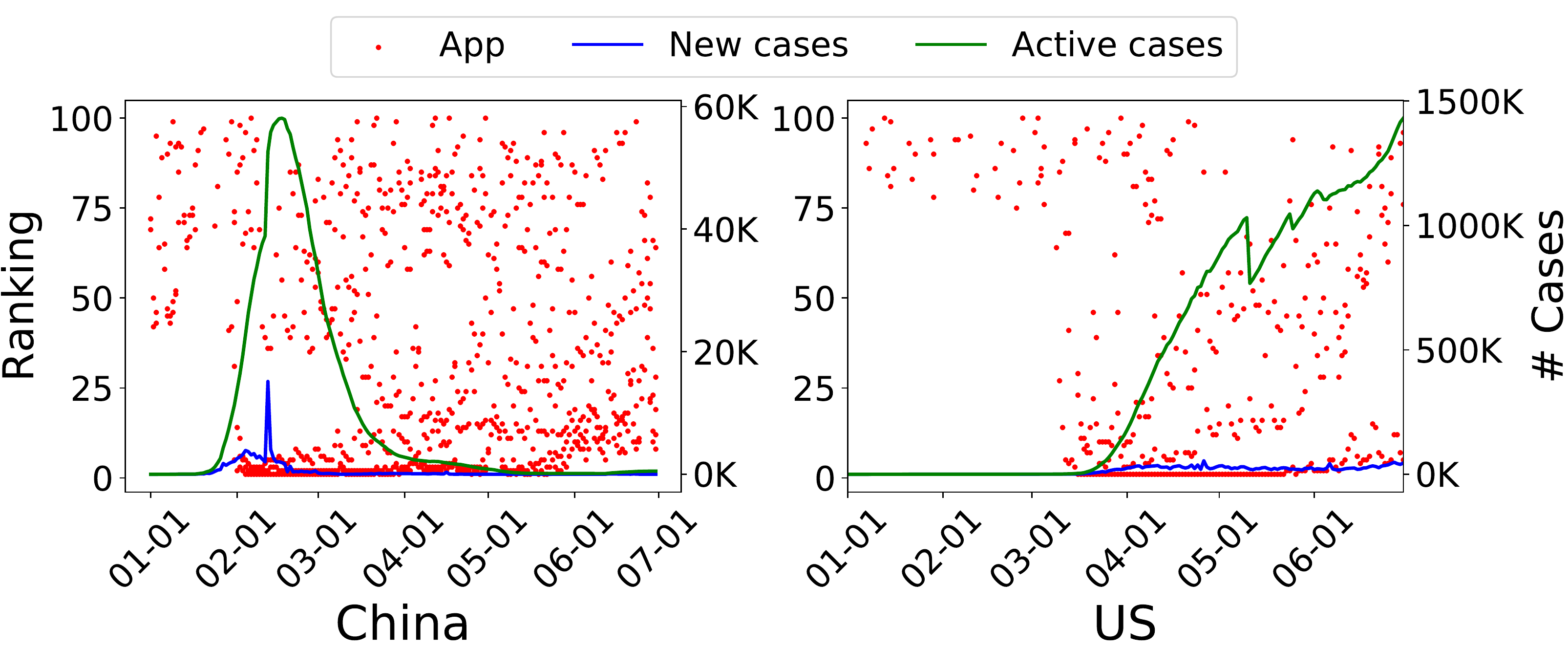}
    }
    \subfigure[Education]{
    \includegraphics[width=0.48\textwidth]{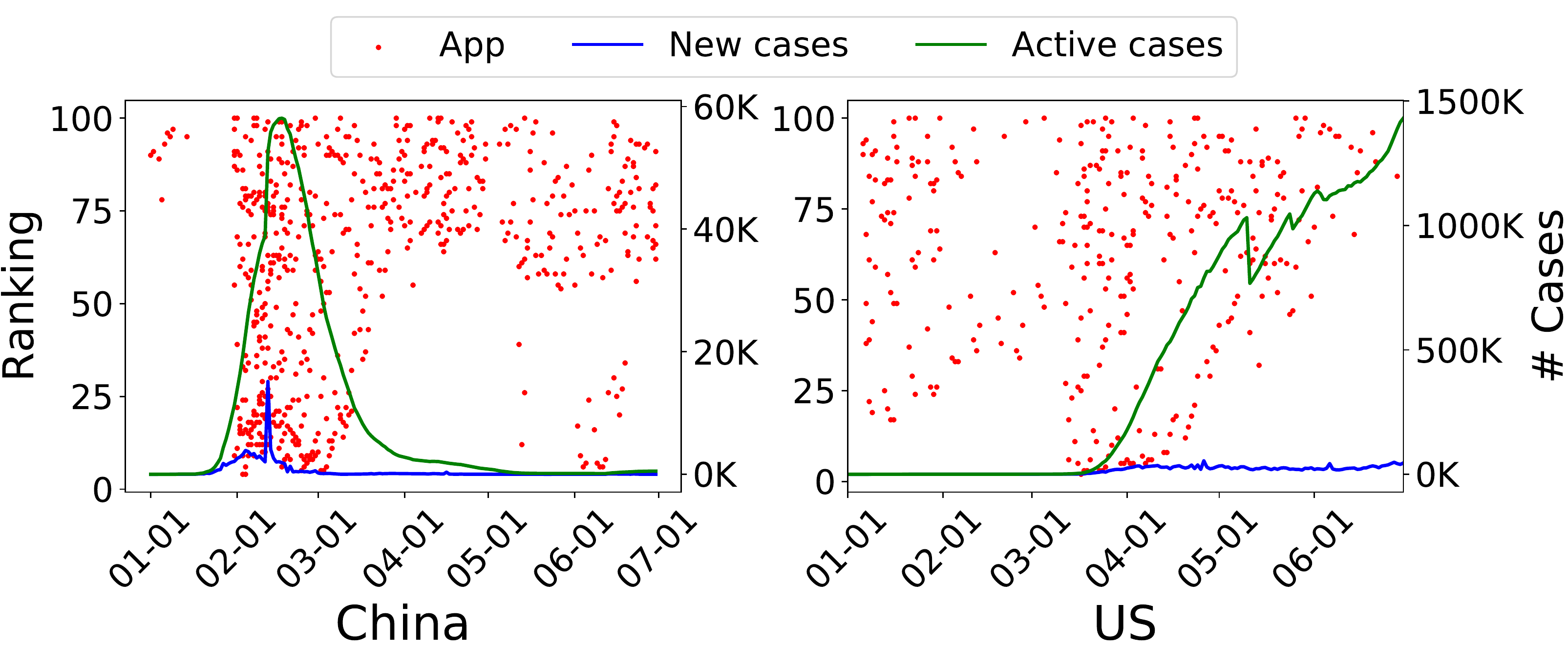}}
    
    \subfigure[Travel]{
    \label{fig:travel}
    \includegraphics[width=0.48\textwidth]{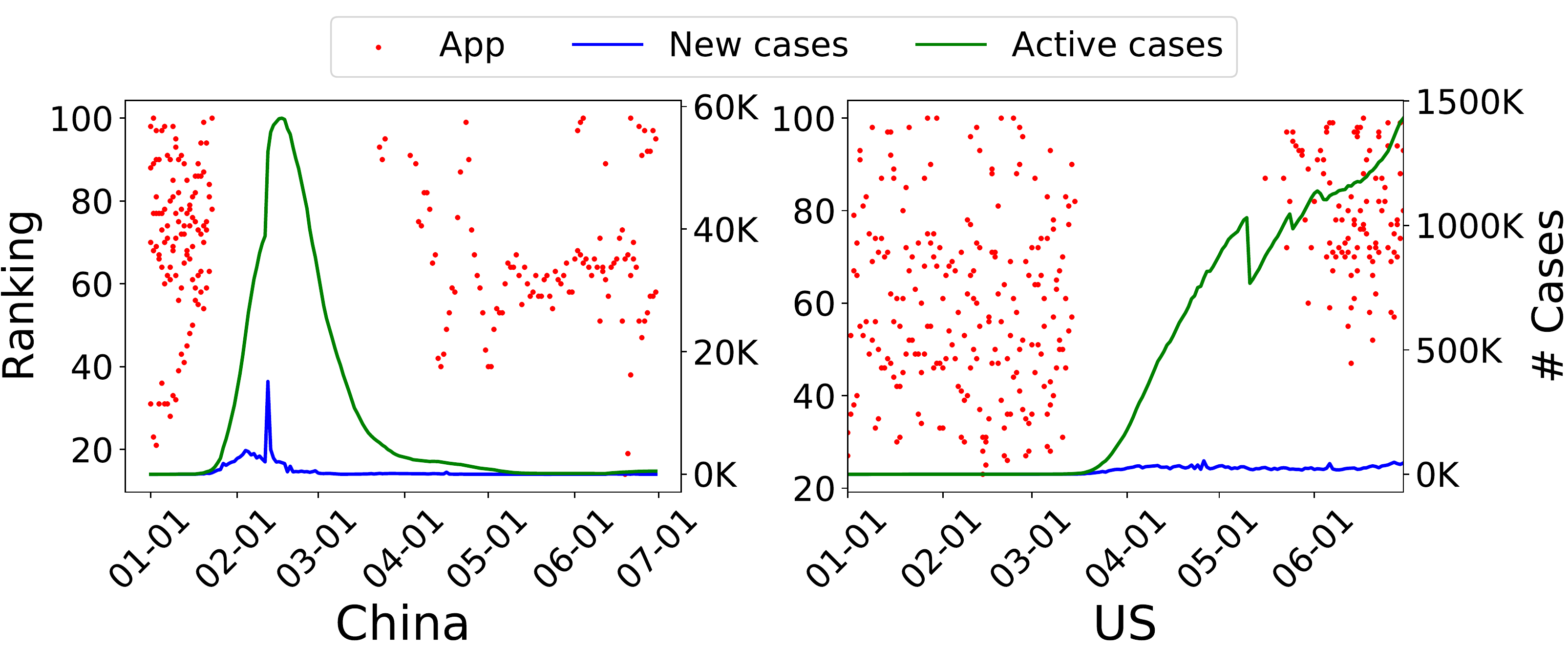}}
    \subfigure[Navigation]{
    \includegraphics[width=0.48\textwidth]{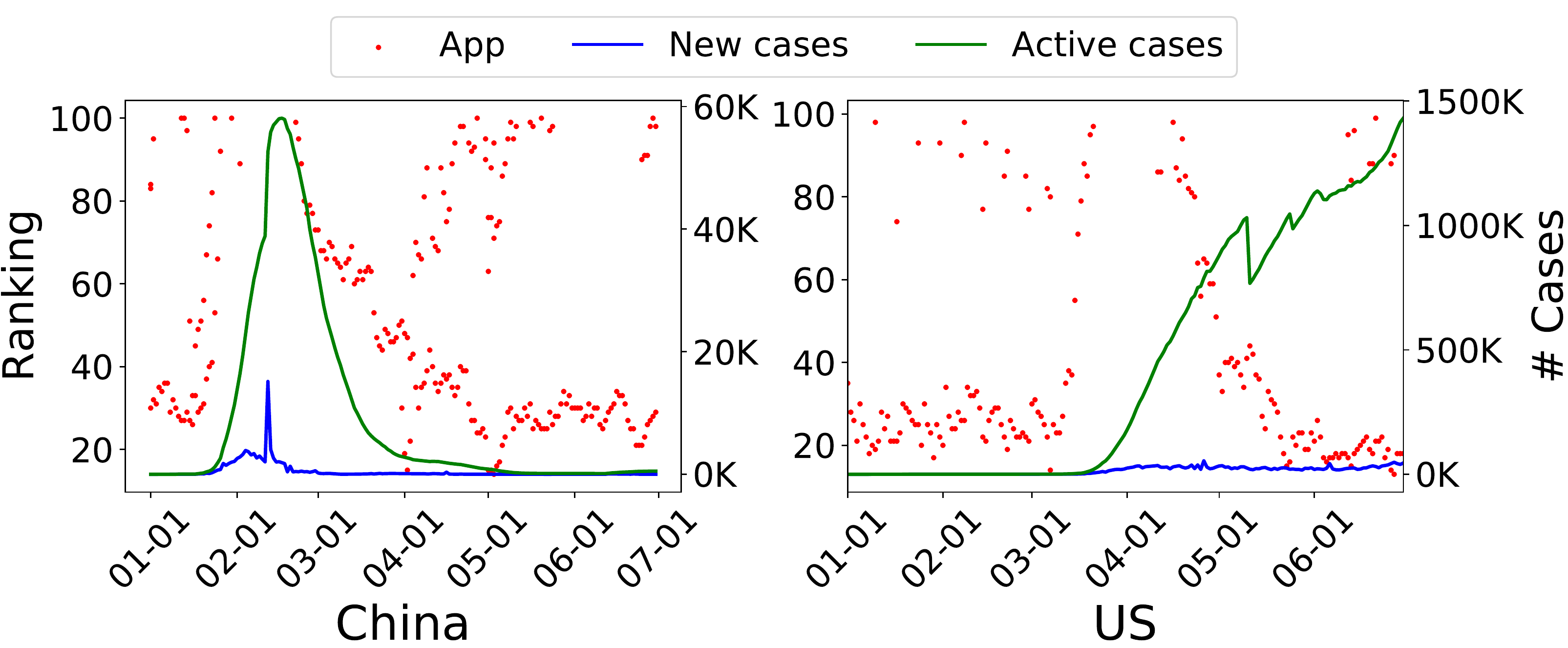}}
    \caption{Evolution of the daily ranking of apps and the number of diagnoses in four categories (Business, Education, Travel and Navigation) in two countries.}
    \label{fig:four-category}
\end{figure*}

For each of the four categories, we plot the overlay of scatter and line charts for both countries in Figure~\ref{fig:four-category}. The scatter represents the ranking of each app in that category and the lines show the number of newly confirmed and active COVID-19 cases per day.
First, we observe a number of \texttt{Business} apps (e.g., Tencent Meetings and Zoom Cloud Meetings) and \texttt{Education} apps (e.g., Tencent Classroom and Google Classroom) jumped to the top of the list.
This started in early February in China and mid-March in the US, just as the COVID-19 began to spread widely in the corresponding countries. This observation is quite straightforward since the outbreak had quickly led to a large-scale shift to remote work for employees and online learning for students. 
Besides, the \texttt{Travel} apps (e.g.,Ctrip and Airbnb) and \texttt{Navigation} apps (e.g., Baidu Maps and Google Maps) show the opposite trend during the same time, with almost all of them falling out of the top 100. This is also reasonable because the mandatory social distancing and quarantines are required for containing virus transmission, which left such apps less desired.
It is important to note that the ranking changes in these four categories starts almost simultaneously with the outbreak, which demonstrates the pandemic's immediate impact. 
In sum, the findings indicate that mobile app ecosystems have had immediate reactions to the COVID-19 outbreak. Regarding app categories, COVID-19 has an effect of enhancing or decreasing the popularity of certain categories of apps.

Comparing the two countries, it is also interesting to note that \texttt{Navigation} and \texttt{Travel} apps in the US rebounded in mid-April and mid-May, respectively, while the US was still in the midst of an increasingly severe pandemic situation. In contrast, the rankings for these two categories in China improved only after the conditions began to improve. To some extent, this may reflect the different attitudes of the two countries in dealing with COVID-19.

\iffalse
\begin{figure*}[htb]
    \centering
    \begin{center}
        \subfigure{
        \includegraphics[width=0.98\textwidth]{Figures/legend.pdf}
        } 
        \vskip -0.15in
        \subfigure[China]{
        \includegraphics[width=0.48\textwidth]{Figures/category-cn.pdf}
        \label{fig:category-cn}}
        \subfigure[US]{
        \includegraphics[width=0.48\textwidth]{Figures/category-us.pdf}
        \label{fig:category-us}}
        \vspace{-0.1in}
        \caption{The distribution of number of apps in each category per day}
        \label{fig:category}
    \end{center}
\end{figure*}
\fi

\subsection{Correlation Analysis}
As mentioned in the above, we employ Pearson Correlation Coefficient to explore the correlations between app popularity and the pandemic situation.
In this section, we provide correlation analysis results at both the overall and category levels, respectively.

\subsubsection{Overall distribution}
\begin{figure}[htb]
    \centering
    
    \includegraphics[width=0.5\textwidth]{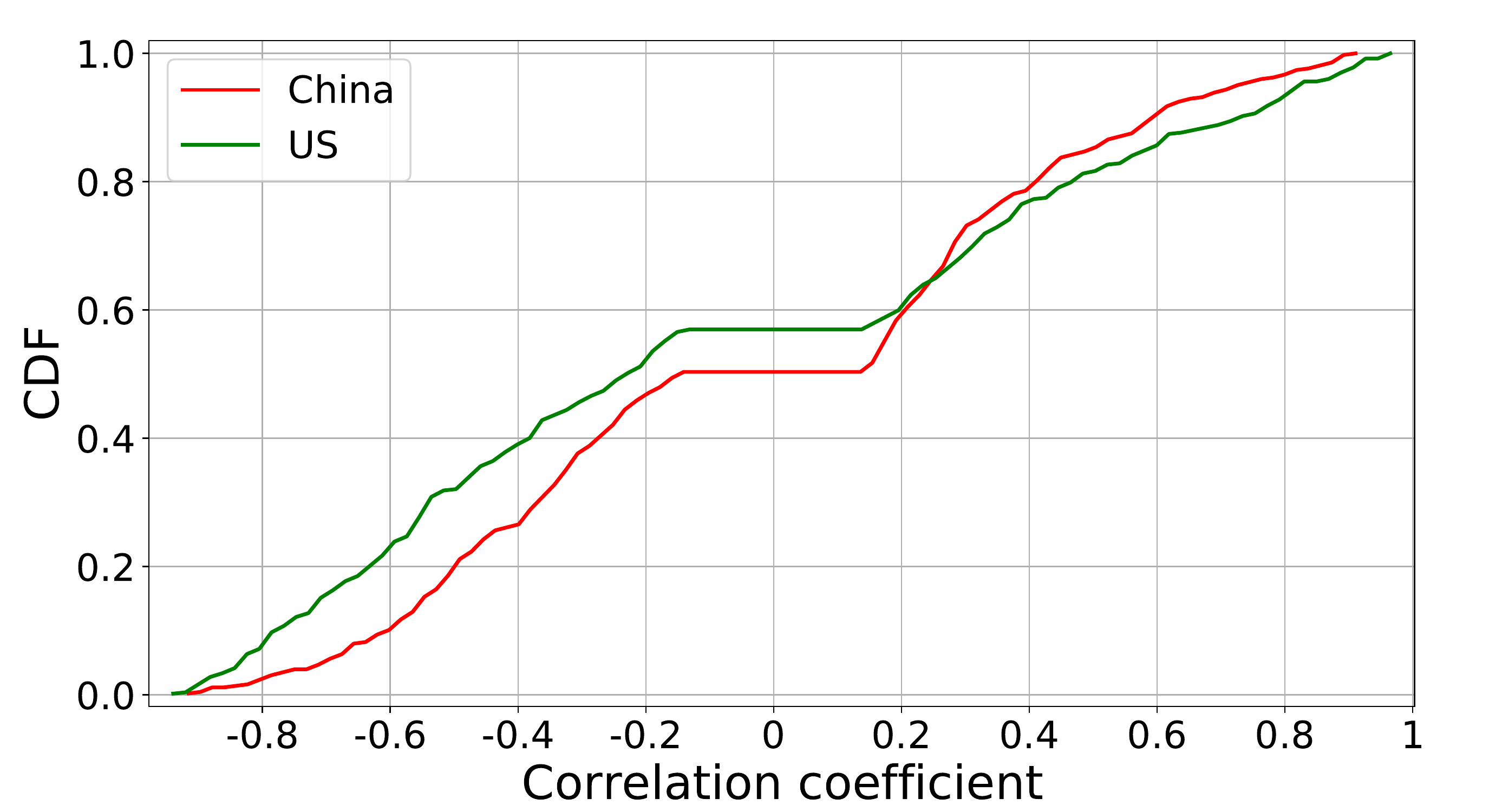}
    \caption{The overall distribution of correlation coefficients}
    \label{fig:corr-overall}
\end{figure}

Figure~\ref{fig:corr-overall} presents the Cumulative Distribution Function (CDF) of the correlation coefficients between app ranking and the number of active confirmed cases in the two countries. 
The trends are very similar in the US and China, with a roughly 50:50 number of positively and negatively correlated apps. Besides, there are indeed a number of apps whose ranking changes have a clear correlation with the pandemic evolution in both countries. Specifically, 27\% of the apps in China have a correlation coefficient less than -0.4 and nearly 40\% in the US. In addition, over 20\% of the apps show a correlation coefficient greater than 0.4 both in China and the US, which manifests many positive correlations. More notably, 23\% of the apps in the US and 10\% in China have correlation coefficients less than -0.6, and nearly 10\% in both countries are greater than 0.6, which can be considered as strong or very strong correlations (see Table~\ref{tab:r-strength}). This observation suggests that many apps' popularity does haves significant correlation with the development of pandemic, positively or negatively.

\subsubsection{Category-level distribution}
\begin{figure}
    \centering
    \includegraphics[width=0.8\textwidth]{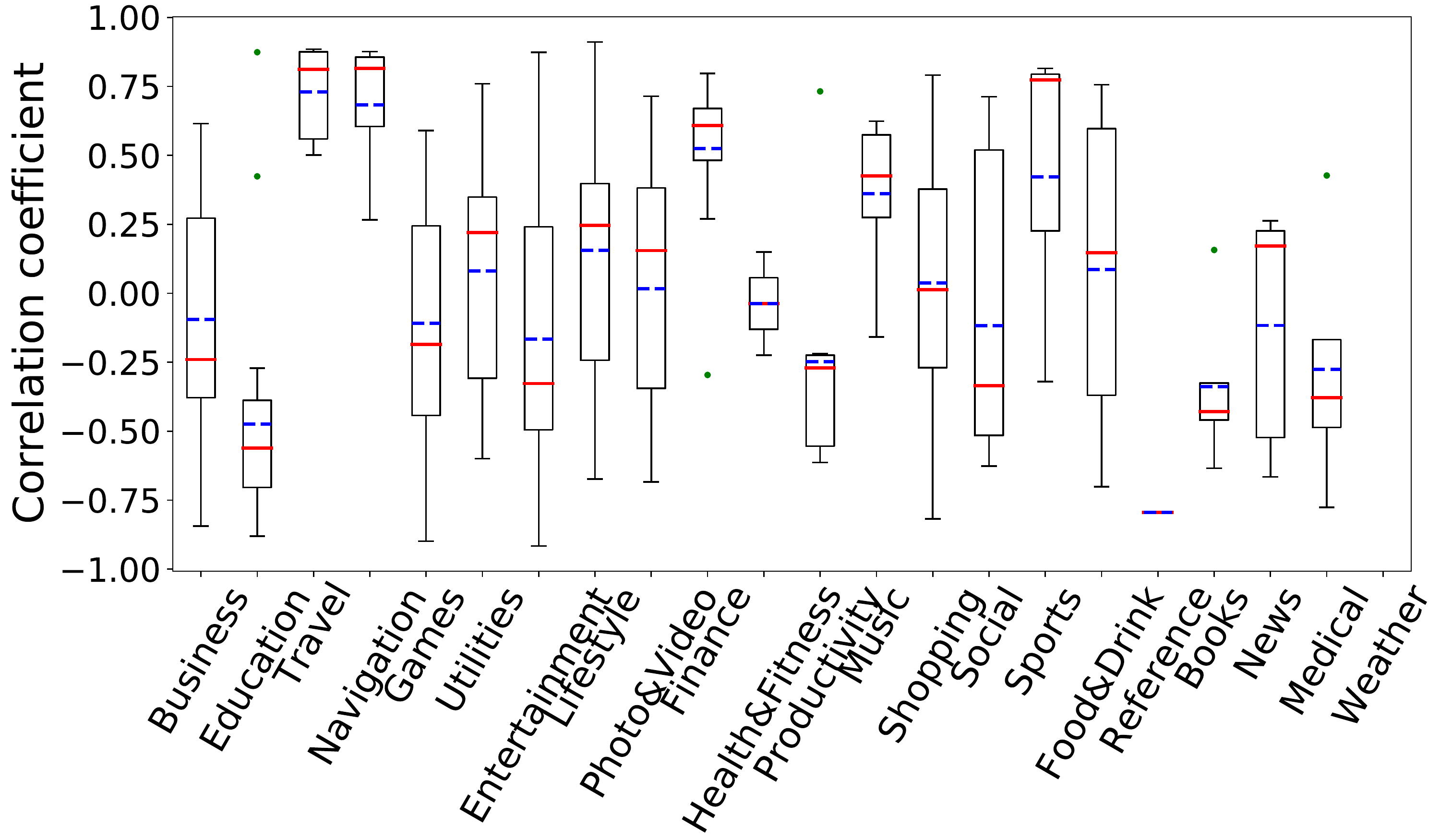}
    \caption{The distribution of correlation coefficients in each category in China}
    \vspace{-0.2in}
    \label{fig:boxplot-cn}
\end{figure}

We further study the distribution of correlations at the category level.
Figure~\ref{fig:boxplot-cn} shows the boxplot of correlation coefficients for each category in \emph{China}. The red line in the box indicates the median, while the blue line represents the mean. First, we observe that the two most prominent categories are \texttt{Travel} and \texttt{Navigation}, with all correlation coefficients being positive and a median greater than 0.75. This indicates that there is a significant positive correlation between pandemic situation and the ranking of apps in these two categories, i.e., the ranking decreases as the pandemic situation worsens and increases when the situation improves (as measured by case rate).

Moreover, most apps in the \texttt{Education} category have negative correlation coefficients with a median of about -0.6, suggesting that \texttt{Education} apps are in general negatively correlated with the pandemic situation.
The findings are consistent with the evolution of the popularity of these three categories, as previously discussed. 
However, for the \texttt{Business} category, there seems to be no significant negative correlation discernible. A possible explanation is the long term popularity for most \texttt{Business} apps after the outbreak, as we can observe in Figure~\ref{fig:business}. There are a number of \texttt{Business} apps that jumped up the rankings during the outbreak and have been leading ever since, even though the number of active cases has dwindled to almost nothing. As a result, there is no significant positive or negative correlation in the later phases, thereby undermining the overall correlation coefficient. We posit that we can examine it in more detail by splitting it into two stages, which we will discuss later.
In general, although most of the categories do not show a significant positive or negative correlation, there are certain categories that do, such as \texttt{Travel} and \texttt{Navigation} showing a strong positive correlation and \texttt{Education} presenting a strong negative correlation.

\begin{figure}
    \centering
    \includegraphics[width=0.8\textwidth]{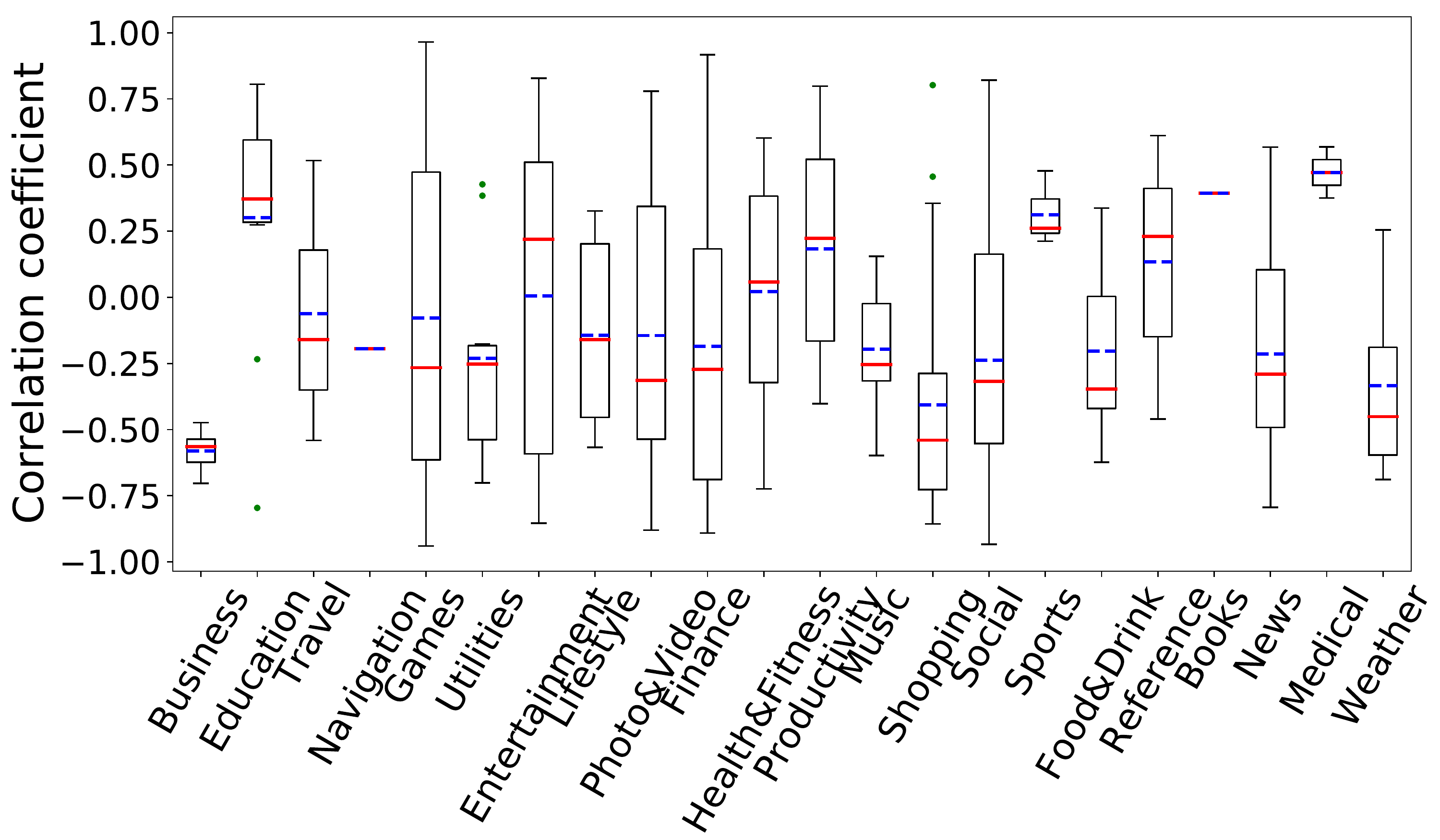}
    \caption{The distribution of correlation coefficients in each category in US}
    \vspace{-0.2in}
    \label{fig:boxplot-us}
\end{figure}

The situation in the \emph{US} is somewhat different. Figure~\ref{fig:boxplot-us} presents the boxplot of correlation coefficients for each category in the US. Surprisingly, there seem to be no conspicuous categories showing significant positive or negative correlations with the pandemic. Again focusing on the Travel category, we can see the correlation coefficients range between -0.5 and 0.5, with a mean close to 0, failing to indicate a strong correlation. 

It is not hard to understand when we review Figure~\ref{fig:travel}. Although the \texttt{Travel} apps have a significant drop in ranking at the beginning, it only lasted for about two months and then gradually rebounded back, during which time the number of diagnoses in general kept increasing, thus there is no noticeable correlation between ranking and pandemic during the entire half year. In the same vein, there is hardly any category in which app rankings are generally growing or declining for a extended period, which weakens the correlations. Therefore, in terms of category level, the correlation between app popularity and the pandemic seems to be less significant.

\subsubsection{Summary}
From the overall-level analysis, we can conclude that there are indeed a number of apps that are strongly correlated with the outbreak situation over the six months of the pandemic, positively or negatively, in both countries. In terms of the category level, the vast majority of categories fail to demonstrate strong correlations.
Besides, we can discover some distinctions in the correlations between app rankings and pandemic conditions in China and the US. In China, there are certain categories (e.g., Travel and Navigation) showing significant correlations, while in the US, the correlations are not very strong in any category. This confirms that the impact of COVID-19 on the mobile app ecosystem may differ from country to country.

\section{Temporal Patterns in App Popularity}
\label{sec:perapp}

We next introduce the findings related to the $\mathbf{RQ_2}$.
Up to now, we have identified that the case rates of the outbreak have an immediate impact on app rankings, with a number of apps significantly correlated with the outbreak.
We then delve into the different patterns that each app ranking conforms to, i.e., the different correlations with the pandemic trends. We make efforts to create a taxonomy and classify the apps based on their calculated correlation coefficients, where each category corresponds to a pattern of change in rankings. 
Considering the different trends in the evolution of the pandemic situation in China and the United States over the first six months, i.e., China has experienced a progression from deterioration to improvement, while the US has tended to be more severe. We therefore examine the situation in China in two stages for a more fine-grained comparison.

The first stage was when the pandemic was worsening, and the second stage was when it was improving. The cut-off date is February 17, when the number of new confirmed cases reached its highest.
% \gareth{How was this date decided?}
Besides, since a larger correlation coefficient shows a more significant correlation, we consider introducing a threshold \emph{n}, and regard an app as strongly positively or negatively correlated with the outbreak (overall or in a stage) whose absolute correlation coefficient is greater than \emph{n}. As such, the key challenge is to select an appropriate \emph{n}. Nevertheless, the correlation coefficient represents an effect size and so we can verbally describe the strength of the correlation using the guide that Evans \cite{evans1996straightforward} suggests for the absolute value of \emph{r} in Table~\ref{tab:r-strength}. Thus, we set the threshold to $0.6$. To be specific, we take $r>0.6$ as strong positive correlation (SPC) and $r<-0.6$ as strong negative correlation (SNC).
Under all these scenarios, we create a taxonomy including eight categories for China and two categories for the US.

\begin{table}[htb]
    \centering
    \vspace{-0.15in}
    \caption{A commonly used interpretations of the \emph{r} values}
    \vspace{-0.1in}
    \begin{tabular}{c|c}
    \toprule
     Magnitude of Correlation & Description of Strength \\
     \midrule
     0.01 - 0.19 & Very Weak \\
     0.20 - 0.39 & Weak \\
     0.40 - 0.59 & Moderate \\
     0.60 - 0.79 & Strong \\
     0.80 - 1.00 & Very Strong \\
     \bottomrule
    \end{tabular}
    \label{tab:r-strength}
\end{table}

\subsection{Taxonomy}
\subsubsection{China}
In China, we consider the situation from both an overall and a phased perspective.

(1) Overall-level.
We start with checking the correlation coefficient \emph{r} between the app ranking and the overall period of the pandemic. If \emph{r}>0.6 we classify the app as ``SPC with the overall.'' Similarly, we categorize the app as ``SNC with the overall'' if its \emph{r}<-0.6.

(2) Stage-level.
For the apps without a strong correlation between its ranking and the overall pandemic period, we then examine their correlations with the two stages respectively, i.e., the growth stage of the pandemic and the decline stage of the pandemic. If an app's rank is strongly correlated to only one of the stages, we label it as a one-stage correlation. Concretely, there are four categories including ``SPC with growth stage'', ``SNC with growth stage'', ``SPC with decline stage'' and ``SNC with decline stage.'' 

If an app's ranking is strongly correlated with both stages but in opposite directions, we deem it as two-stages correlation. Specifically, there are two categories including ``SPC with growth stage, SNC with decline stage'' and ``SNC with growth stage, SPC with decline stage.''

\subsubsection{US}
In the US, the situation is relatively straightforward due to the single upward trend of the pandemic, and we can simply focus on two categories, i.e., ``SPC with the overall'' and ``SNC with the overall.''

\subsubsection{Results}
Based on the above, we classify the apps in the two countries separately and the results are shown in Table~\ref{tab:taxonomy}. For China, we have eight categories due to the shifts in the pandemic trend, with the number of apps ranging from 2 to 44. In the US, there are two categories with app quantities of 72 and 116, respectively.

\begin{table}[htb]
    \centering
    \caption{Eight categories in China and two categories in the US}
    \vspace{-0.1in}
    \resizebox{0.99\textwidth}{!}{
    \begin{tabular}{|l|l|l|c|}
    \toprule
    \multicolumn{1}{|l|}{} & \textbf{Category}  & \textbf{Ranking/Popularity Change} & \textbf{\# Apps} \\ 
    \midrule
    \multirow{2}{*}{Overall}  & SPC with the overall & The app ranking/popularity decreases significantly during the growth stage & 41\\ & & of the pandemic and increases significantly during the decline stage. &  \\ \cline{2-4}
    & SNC with the overall & The app ranking/popularity increases significantly during the growth stage & 44\\ & & of the pandemic and decreases significantly during the decline stage. &  \\ \hline
    \multicolumn{1}{|l|}{\multirow{4}{*}{One-stage}} 
    & SPC with the growth stage  & The app ranking/popularity decreases significantly during the growth stage & 20\\&&of the pandemic but no significant increase or decrease in the decline stage. & \\ \cline{2-4}
    & SNC with the growth stage & The app ranking/popularity increases significantly during the growth stage & 50\\ &&of the pandemic but no significant increase or decrease in the decline stage. &  \\ \cline{2-4}
    & SPC with the decline stage & The app ranking/popularity increases significantly during the decline stage & 22\\ &&of the pandemic but no significant increase or decrease in the growth stage. & \\ \cline{2-4}
    & SNC with the decline stage & The app ranking/popularity decreases significantly during the decline stage & 19\\&&of the pandemic but no significant increase or decrease in the growth stage. & \\ \hline
    \multicolumn{1}{|l|}{\multirow{2}{*}{two-stages}} 
    & SPC with growth stage,&The app ranking/popularity decreases significantly during the growth stage & 7\\ & and SNC with decline stage &of the pandemic and decreases significantly during the decline stage. & \\ \cline{2-4}
    & SNC with growth stage,& The app ranking/popularity increases significantly during the growth stage & 2\\ & and SPC with decline stage &of the pandemic and increases significantly during the decline stage. & \\
    \bottomrule
    \hline
    \multicolumn{1}{|l|}{
    \multirow{2}{*}{US}} &  SPC with the overall & The app ranking/popularity decreases significantly during the period. & 72 \\ \cline{2-4}
                         &  SNC with the overall & The app ranking/popularity increases significantly during the period. & 116 \\
    \bottomrule
    \end{tabular}}
    \label{tab:taxonomy}
\end{table}

\subsection{Per Group Analysis}
As aforementioned, each of the four groups symbolize a pattern of how an app's ranking changes in response to the outbreak. We next deep dive into each group.

\subsubsection{SPC with the overall in China.}
\begin{figure}[htb]
    \centering
    \subfigure[Dianping]{
    \includegraphics[width=0.4\textwidth]{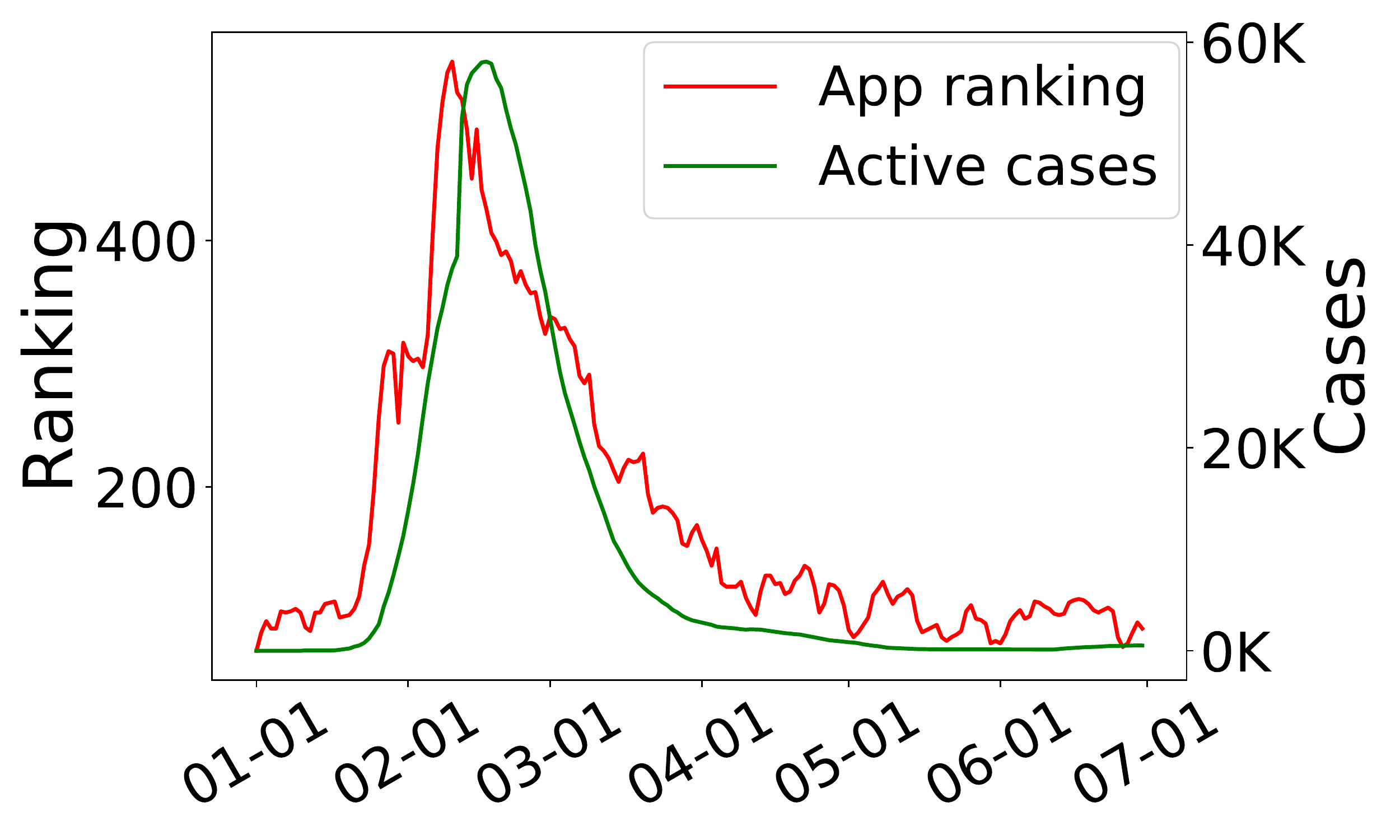}}
    \quad
    \subfigure[TikTak Travel]{
    \includegraphics[width=0.4\textwidth]{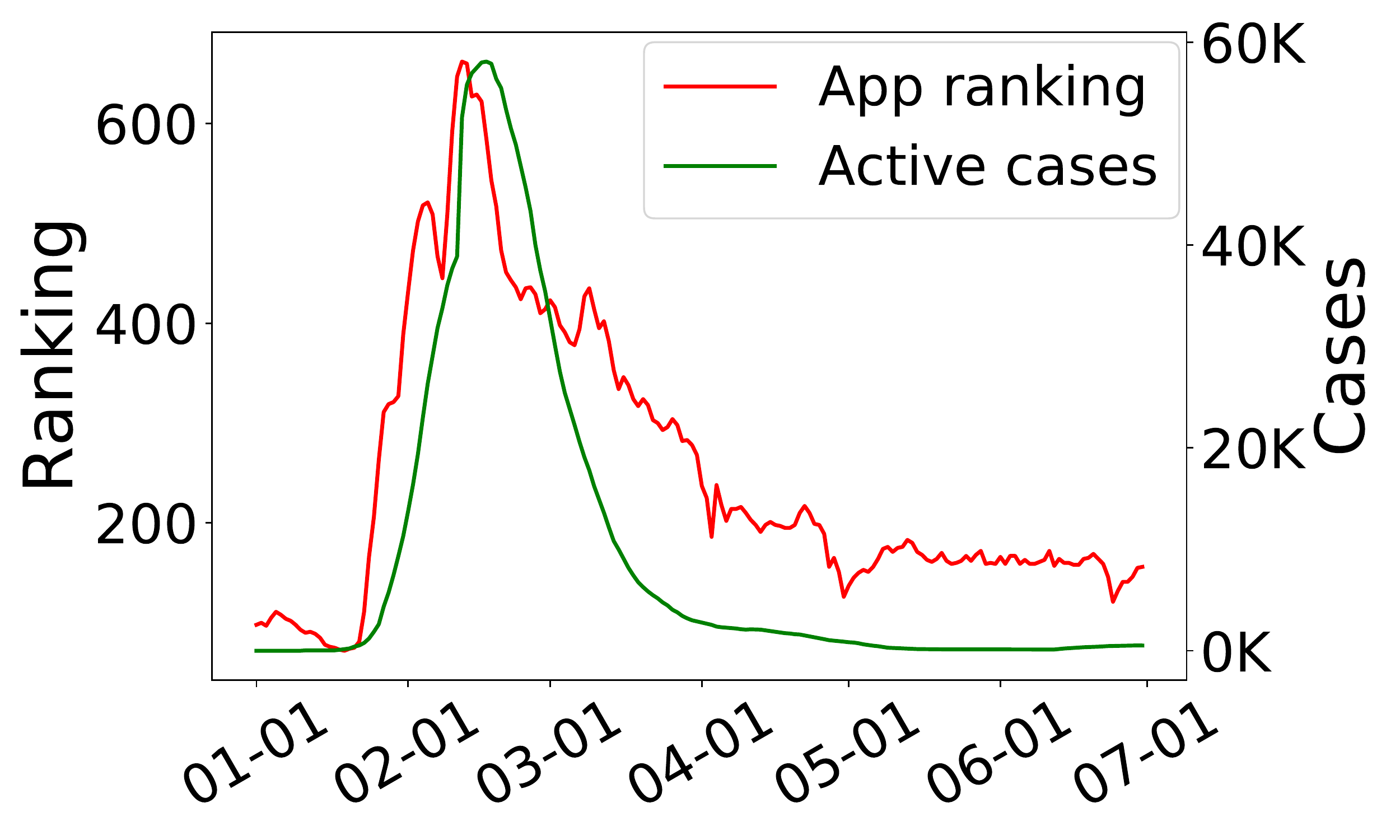}}
    \vspace{-0.15in}
    \caption{Two examples for SPC with the overall in China.}
    \label{fig:c1-example}
\end{figure}

There are 41 apps in China whose app rankings are strongly positively related to the pandemic situation, i.e., they suffered a regression in ranking when the pandemic got worse (measured by case rate). Subsequently, their rankings rebounded as the situation recovered. 
The app with the strongest correlation (0.911) is \textit{Dianping}, a leading local lifestyle information and trading platform in China, providing users with information services such as merchant information, consumer reviews and offers, as well as transaction services including group buying, restaurant reservations, takeaway and e-membership card. It is followed by \textit{TikTak Travel} (0.885), \textit{Ctrip Travel} (0.882) and \textit{Tencent Maps} (0.876), all of which cater to services like travel and navigation. Unsurprisingly, most of the apps in this category are closely tied to outdoor activities. Given the strict quarantine policy during the outbreak in China, it is no wonder that the rankings of this kind of apps undergo such a shift. The trend comparisons for two representative apps are visualized in Figure~\ref{fig:c1-example}. The red line shows the ranking of the app each day and the green line presents the number of active confirmed cases per day. We can observe that their trends are almost identical.

\subsubsection{SNC with the overall in China.}
\begin{figure}[htb]
    \centering
    \subfigure[Diandu]{
    \includegraphics[width=0.4\textwidth]{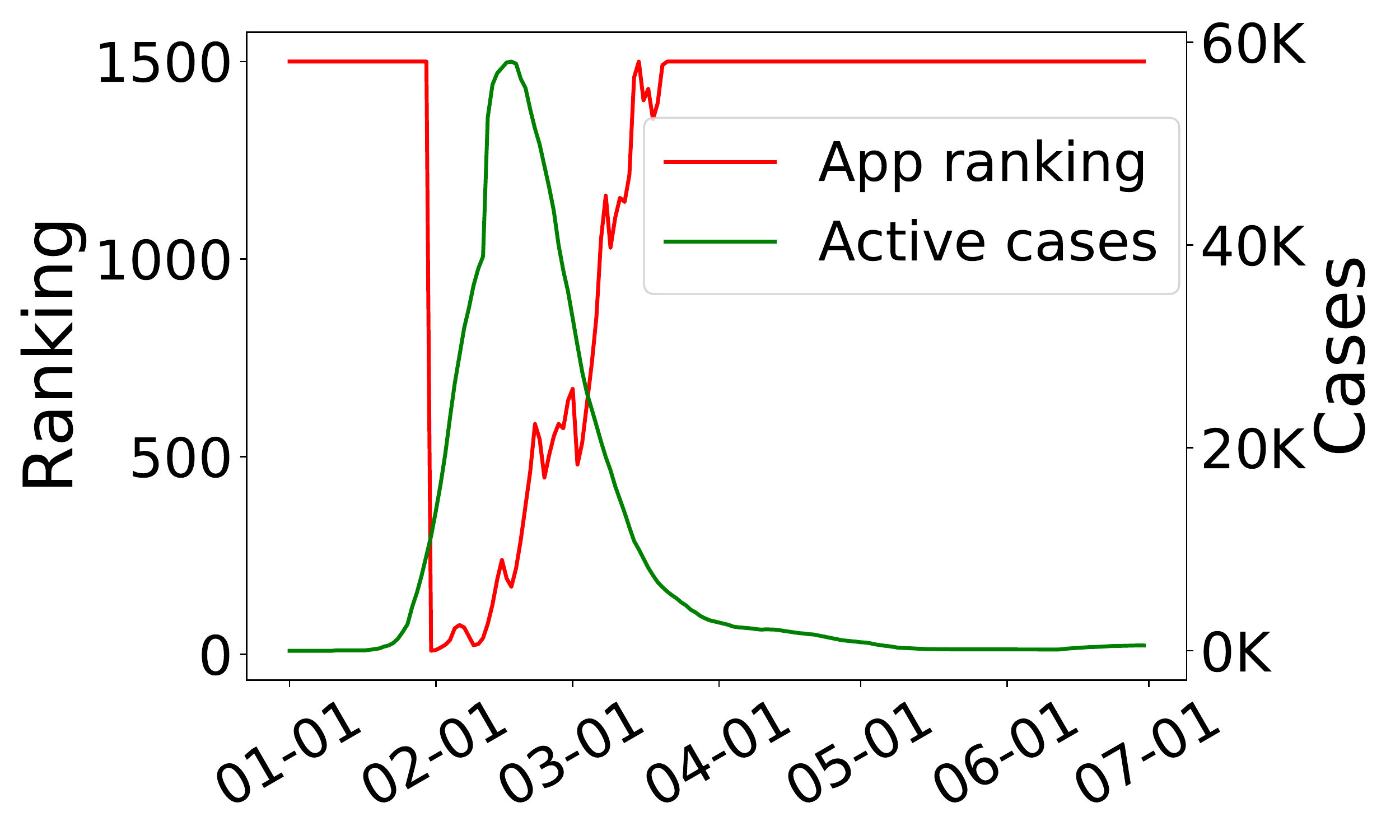}}
    \quad
    \subfigure[Yong Hui Buy Food]{
    \includegraphics[width=0.4\textwidth]{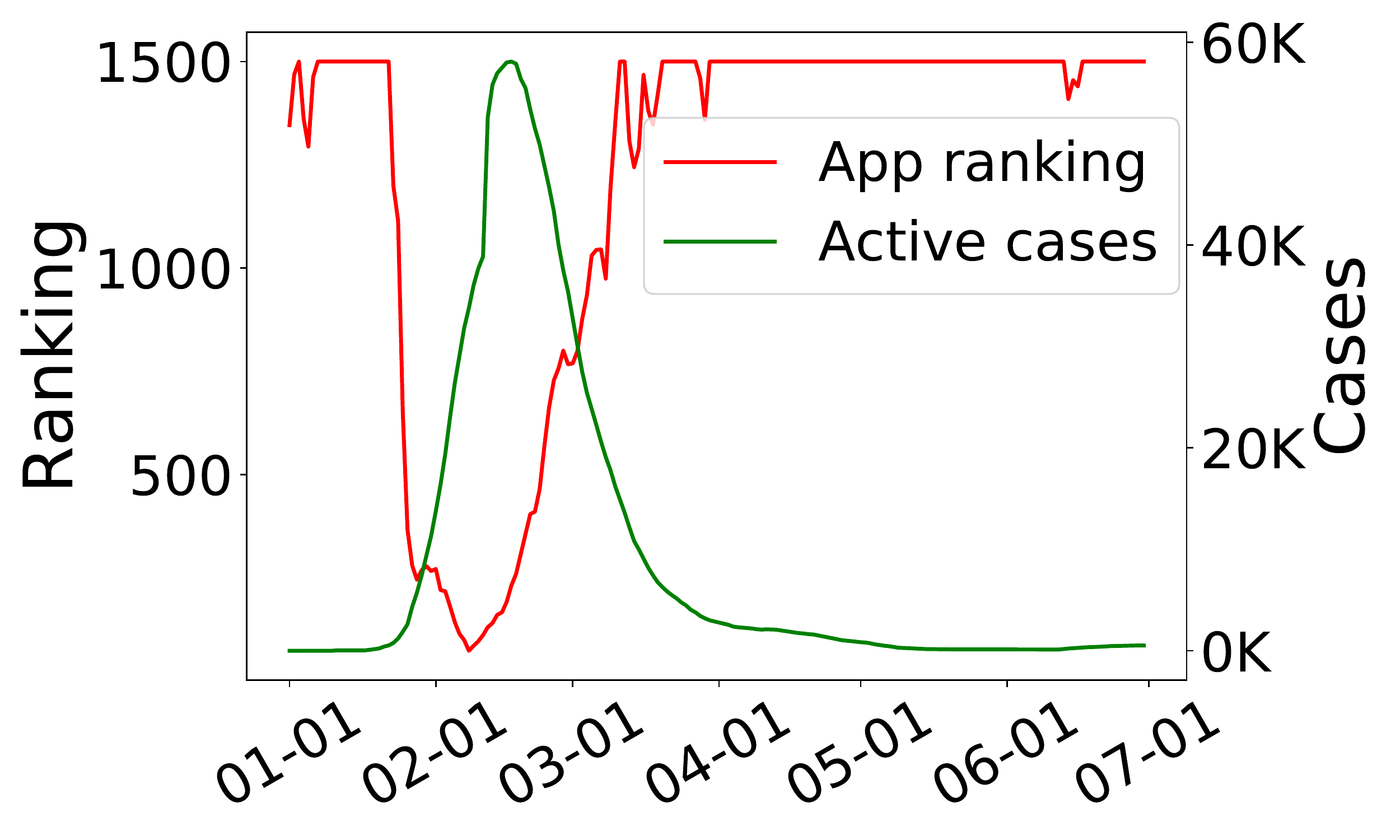}}
    \vspace{-0.15in}
    \caption{Two examples for SNC with the overall in China.}
    \label{fig:c2-example}
\end{figure}

The situation in this category is opposite to the previous one. There is a strong negative correlation between the app ranking and the pandemic case rate trends for the 44 apps in this category. This means that the outbreak has contributed to the growth of their rankings. The subsequent recovery has therefore resulted in a drop in the rankings. 
For context, Figure~\ref{fig:c2-example} presents two example apps with these properties. We see opposite trends in the app ranking vs.\ case rate. Note that since the Apple Store's app ranking list offers up to 1,500 slots, for rankings that fall out of the top 1,500, we replace them with 1,500 in the chart.

We highly four prominent examples that exhibit these behaviors:  
1) \textit{Happy Disappear}, a triple-elimination casual game with the strongest Pearson correlation coefficient of -0.916, which may be a good choice for people to pass time during the lockdown. 
2) \textit{Diandu} (-0.88), an educational app for primary and secondary school language, mathematics, and English learning, which is helpful for tutoring children during the extended period of home study. 
3) \textit{Yong Hui Buy Food} (-0.818), an online service platform that provides consumers with fresh food and other grocery products through home delivery services, which alleviates the inconvenience of going out to buy food during quarantine. 
4) \textit{Tinker Medicine} (-0.776), a pharmaceutical and health product that helps pharmacies provide convenient services to the public and provides free delivery services for users who place orders through the app.
Overall, we observe that most apps in this category have a positive effect on people's lives in some way during the pandemic.

\subsubsection{SPC with the growth stage in China.} 
\begin{figure}[htb]
    \centering
    \subfigure[Eastern Airlines]{
    \includegraphics[width=0.4\textwidth]{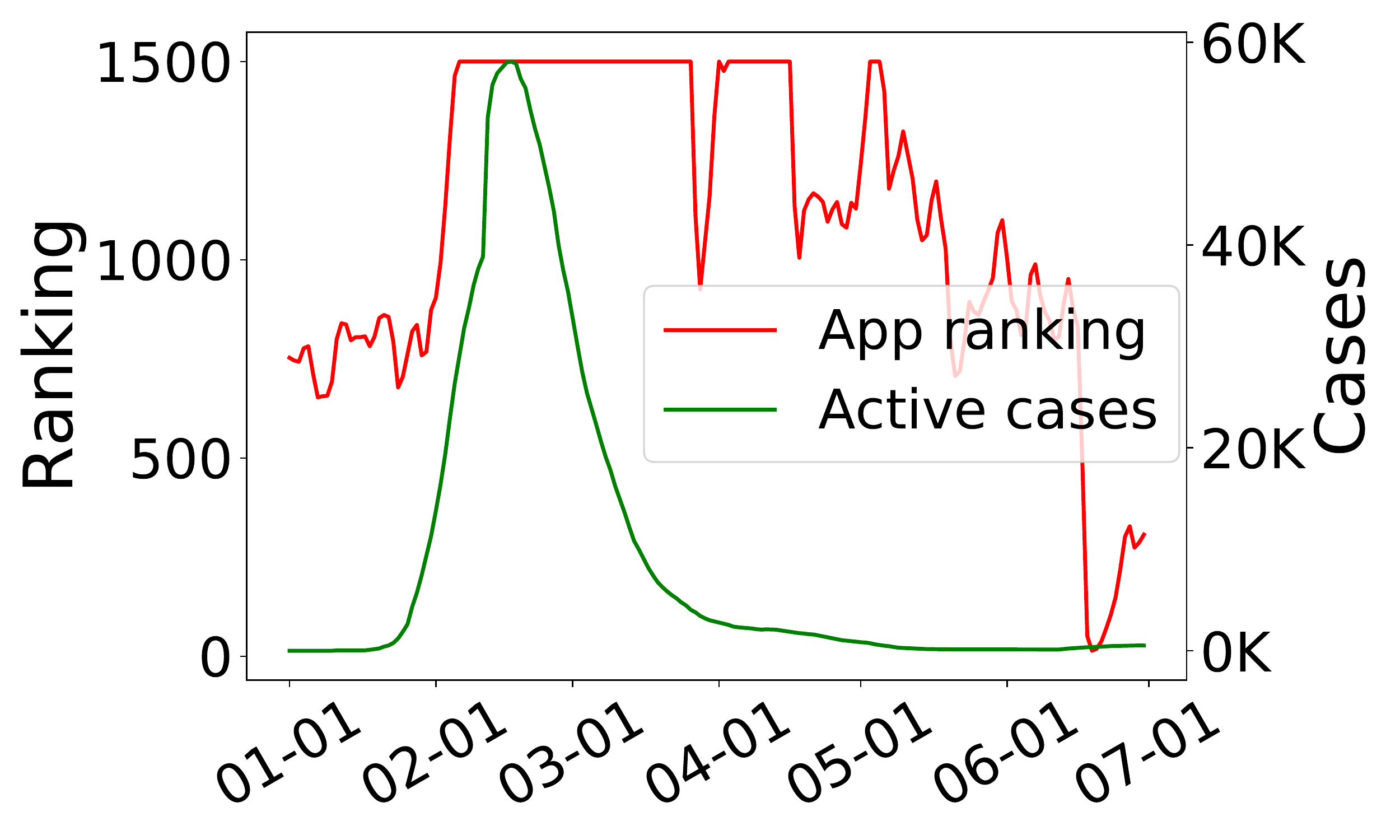}}
    \quad
    \subfigure[Damek]{
    \includegraphics[width=0.4\textwidth]{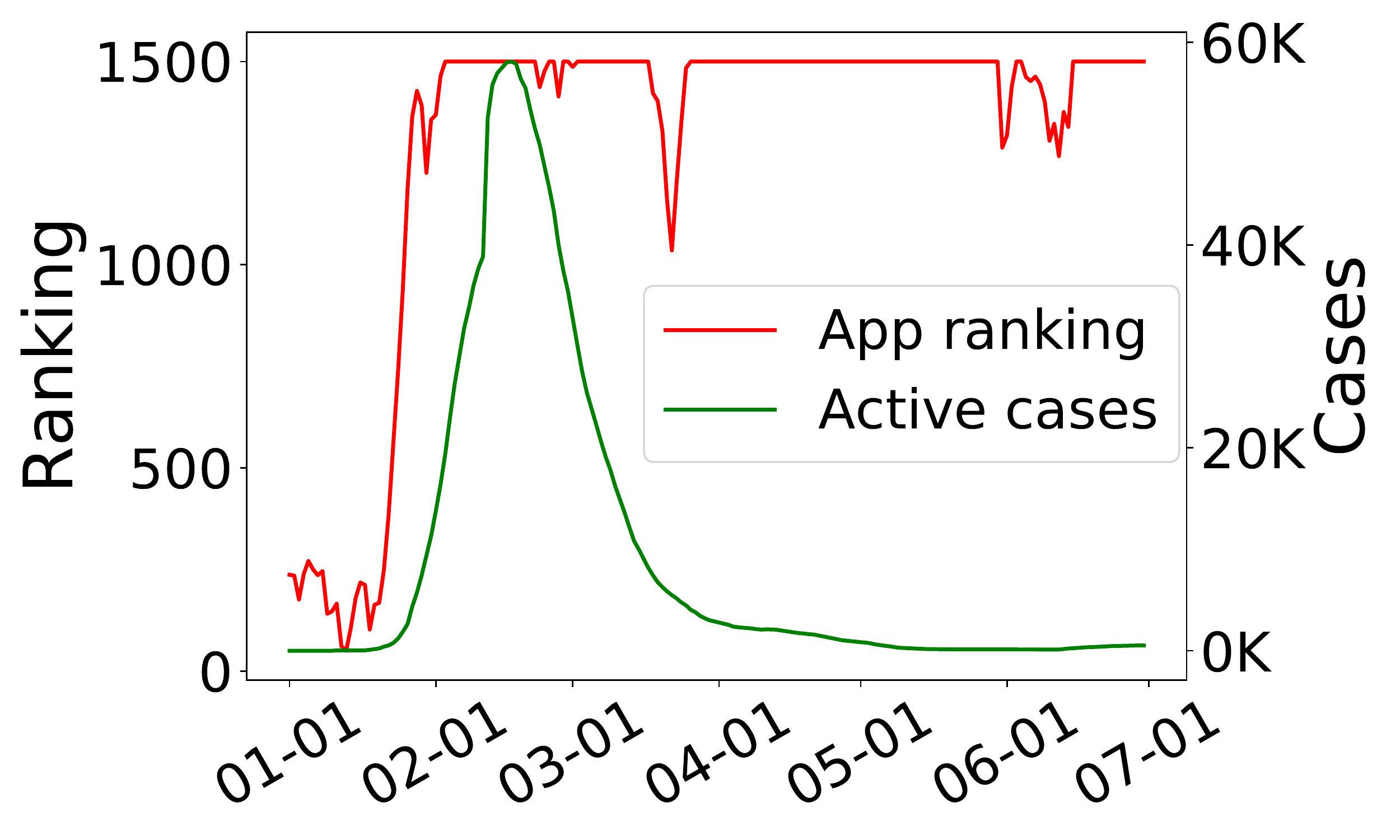}}
    \vspace{-0.15in}
    \caption{Two examples for SPC with the growth stage in China.}
    \label{fig:c3-example}
\end{figure}

In this category, the app ranking is strongly positively correlated only with the growth stage of the pandemic. In other words, as the situation gets worse, the app ranking drops; yet it does not show a definitive upward or downward trend as the situation recovers. For example, there are two typical apps shown in Figure~\ref{fig:c3-example}.
One is called \textit{Eastern Airlines} that aims to provide safe and convenient ticketing and travel experiences. The other is \textit{Damek}, a comprehensive live entertainment ticket marketing platform in China, covering concerts, dramas, musicals, sporting events, etc. Both of them have a high correlation coefficient between app ranking and active cases in the pandemic growth stage (0.933 and 0.772, respectively). However, such correlations are not that strong during the decline stage. This meas there are some apps that are adversely affected by the pandemic but do not recover in tandem with the pandemic.

\subsubsection{SNC with the growth stage in China.}
Similarly, we investigate the situation where app rankings are only strongly negatively correlated with the \emph{growth} stage of the pandemic, i.e., as the pandemic worsened, the app ranking improved, but there was no clear trend of a growing or declining thereafter. 
We take a look at the following examples that have high correlation coefficients:
1) \textit{Tencent Classroom}  (-0.943) is a professional online education platform that provides teachers with online teaching and students with interactive learning, which has been a great support for digital learning due to the school closures.
2)	\textit{Anhui Things} (-0.93) is a mobile app for Anhui Province Government Services Network designed to provide government services, high-frequency public services and citizen-friendly service matters, which allows citizens to handle some business without leaving home.
3) \textit{Tencent Meetings}  (-0.801) is a video conferencing product that features online meetings, useful for the needs of users working from home. 
This indicates that there are a number of apps that played an important role in the war against the pandemic, and continue to be used. This may indicate longer-term implications in how people live and work after the pandemic. 

\begin{figure}[htb]
    \centering
    \subfigure[Tencent Classroom]{
    \includegraphics[width=0.4\textwidth]{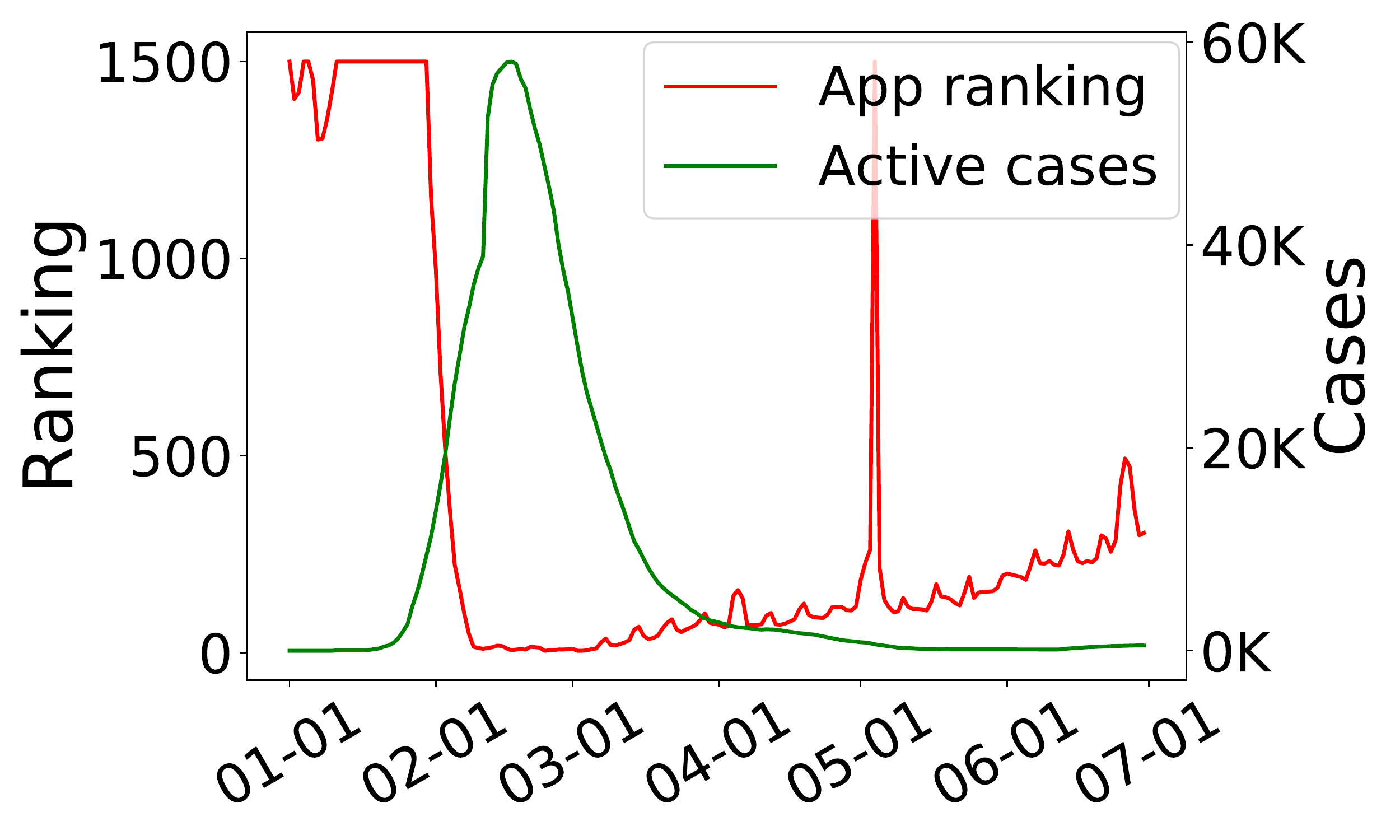}}
    \quad
    \subfigure[Tencent Meetings]{
    \includegraphics[width=0.4\textwidth]{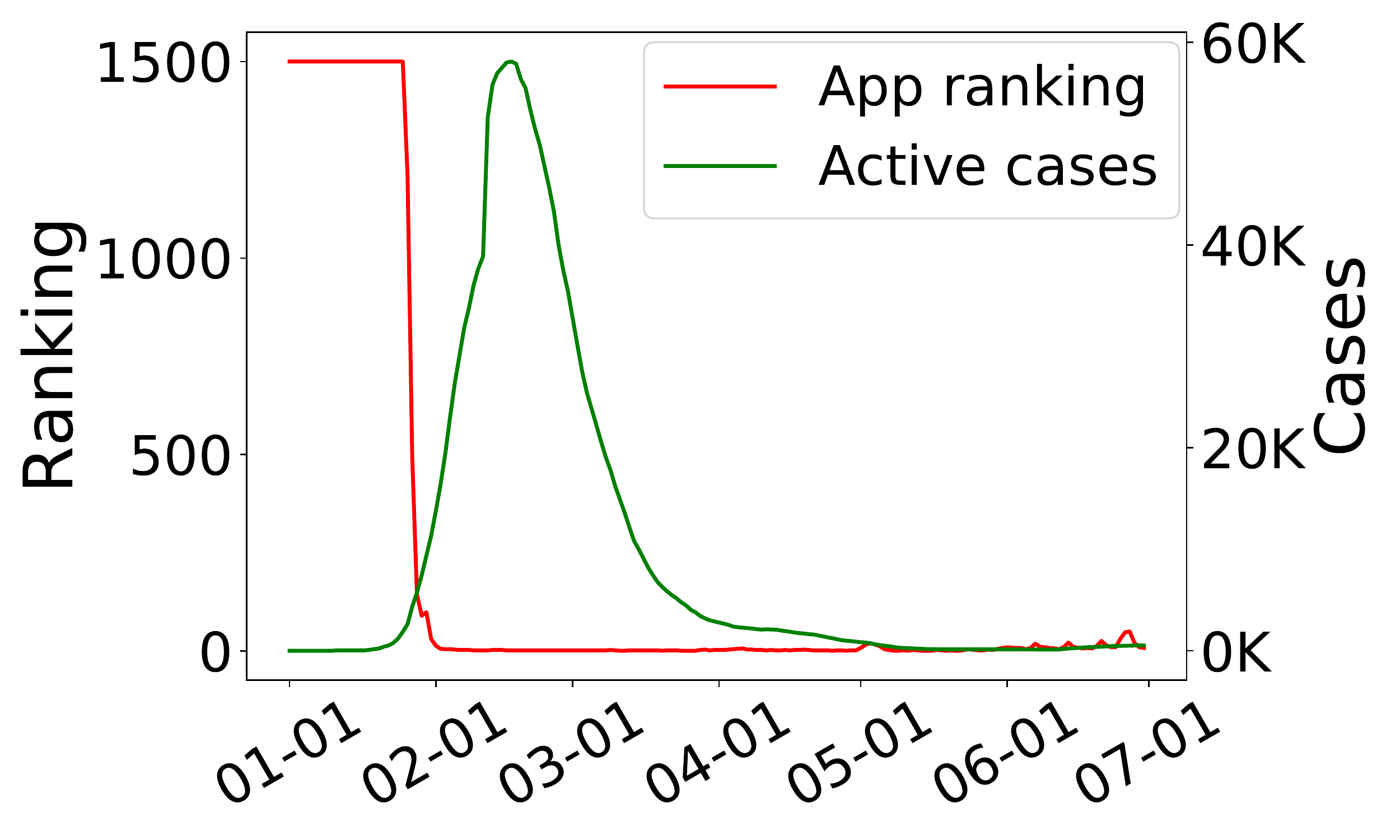}}
    \vspace{-0.15in}
    \caption{Two examples for SNC with the growth stage in China.}
    \label{fig:c4-example}
\end{figure}

\subsubsection{SPC with decline stage in China.}
\begin{figure}[htb]
    \centering
    \subfigure[Shell Finder]{
    \includegraphics[width=0.4\textwidth]{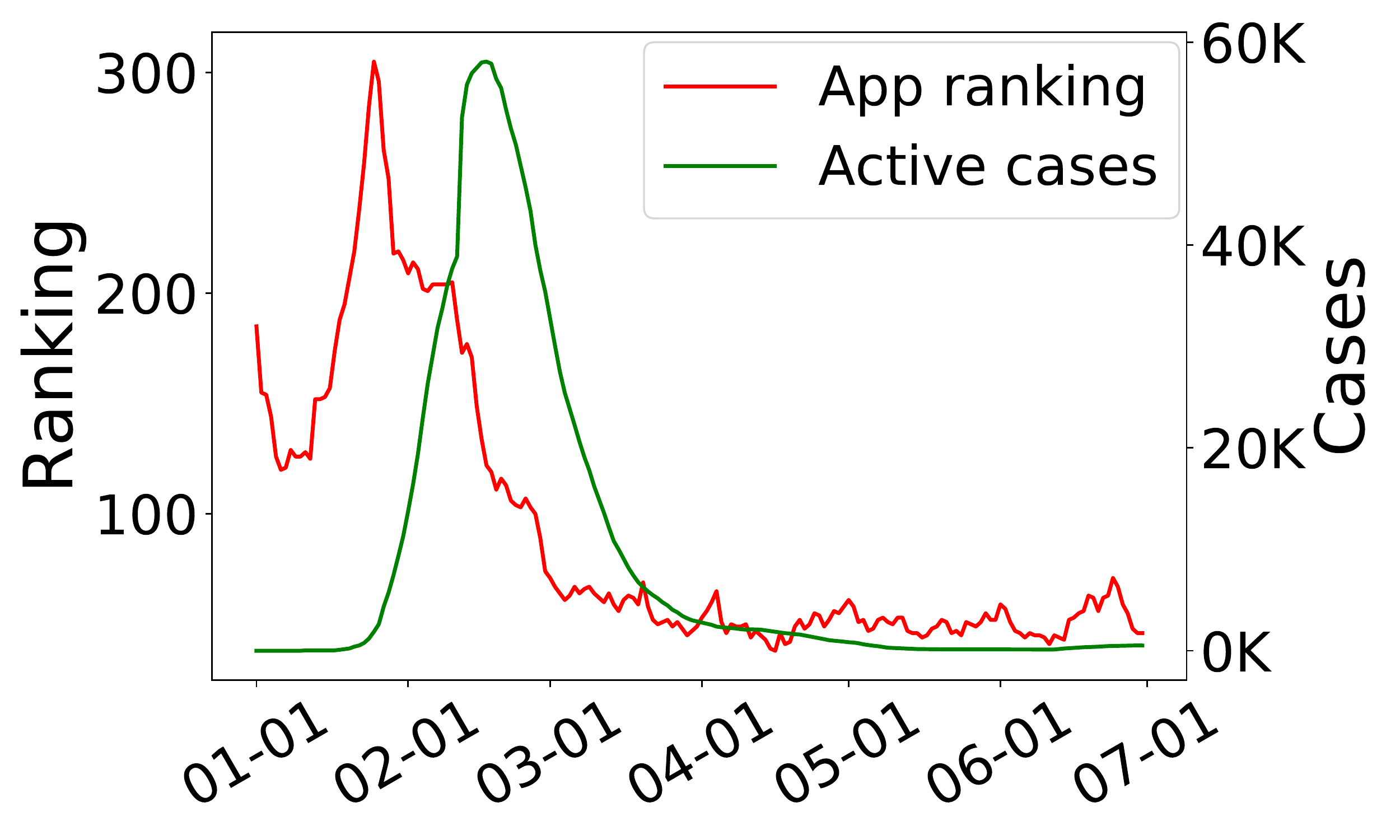}}
    \quad
    \subfigure[Traffic Management 12123]{
    \includegraphics[width=0.4\textwidth]{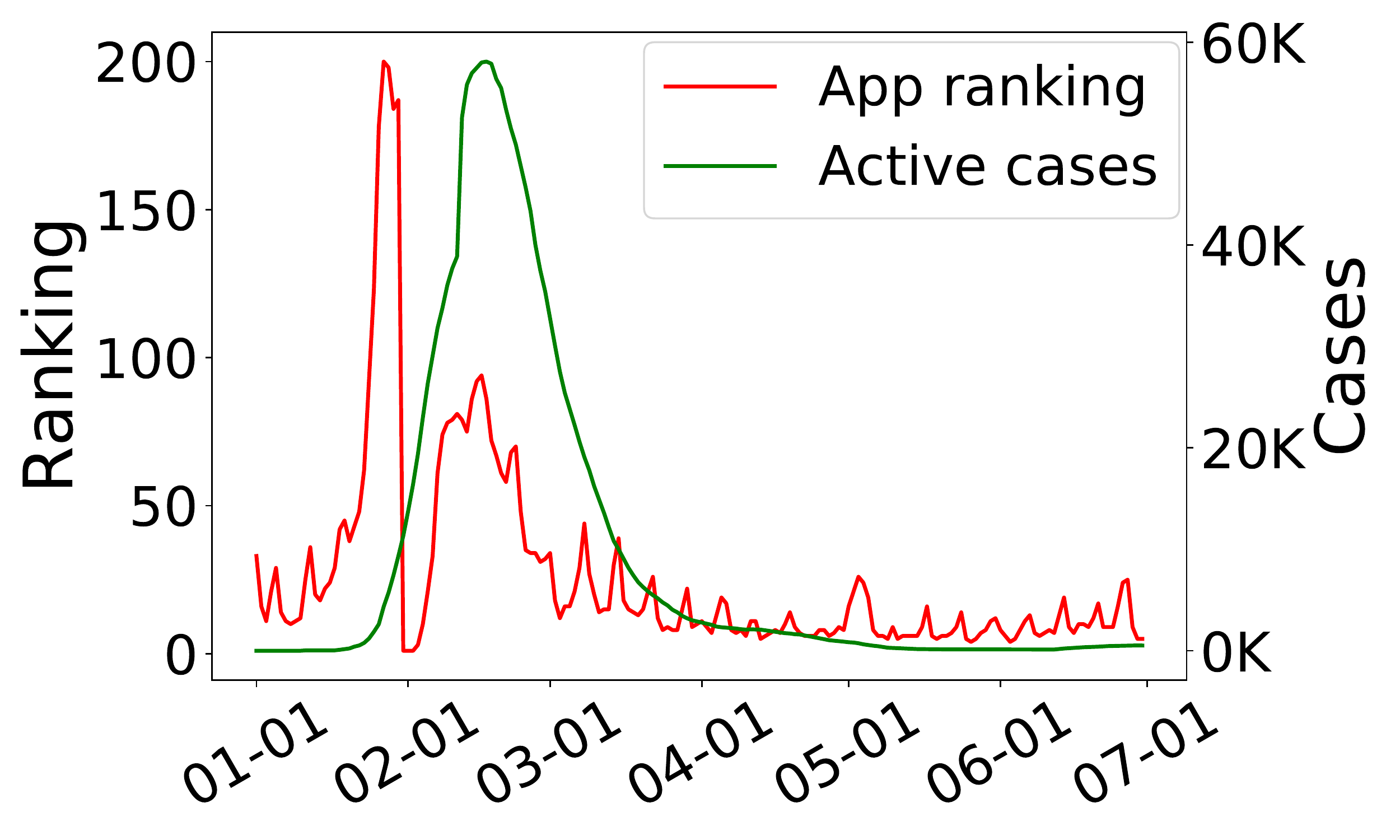}}
    \vspace{-0.15in}
    \caption{Two examples for SPC with the decline stage in China.}
    \label{fig:c5-example}
\end{figure}

In this group, the correlation between app ranking and the pandemic case rate is not significant in the first stage of the pandemic, but as the situation turns better, the app ranking rises. 
For example, the app with the highest correlation coefficient (0.924) in the decline stage is \textit{Shell Finder}, a home searching platform featuring comprehensive and real property information as well as industry innovations such as VR viewings, home valuation, and intelligent recommendations.
In addition, \textit{Traffic Management 12123} (0.881) is the official client of the Internet traffic safety management platform, providing a full range of traffic services and reservation of motor vehicle, driver's license and illegal processing business, etc. As shown in Figure~\ref{fig:c5-example}, we can observe that the rankings of these apps were only slightly affected in the early days of the pandemic.
This reflects the fact that as the situation improves and people's lives gradually return to normal, some apps are coming back to the forefront as soon as possible.

\subsubsection{SNC with decline stage in China.}
\begin{figure}[htb]
    \centering
    \subfigure[Fast View]{
    \label{fig:fastview}
    \includegraphics[width=0.4\textwidth]{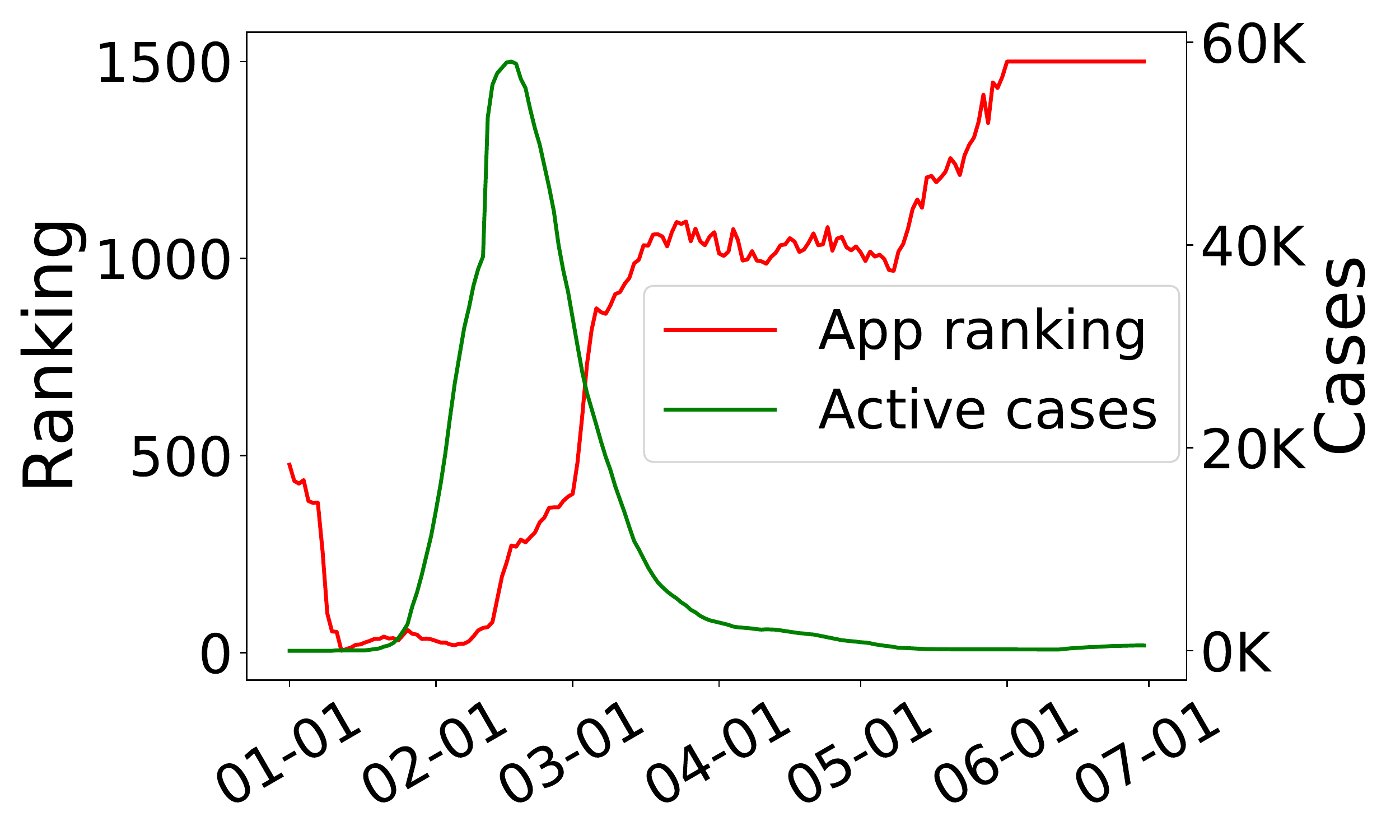}}
    \quad
    \subfigure[Doudizhu]{
    \label{fig:doudizhu}
    \includegraphics[width=0.4\textwidth]{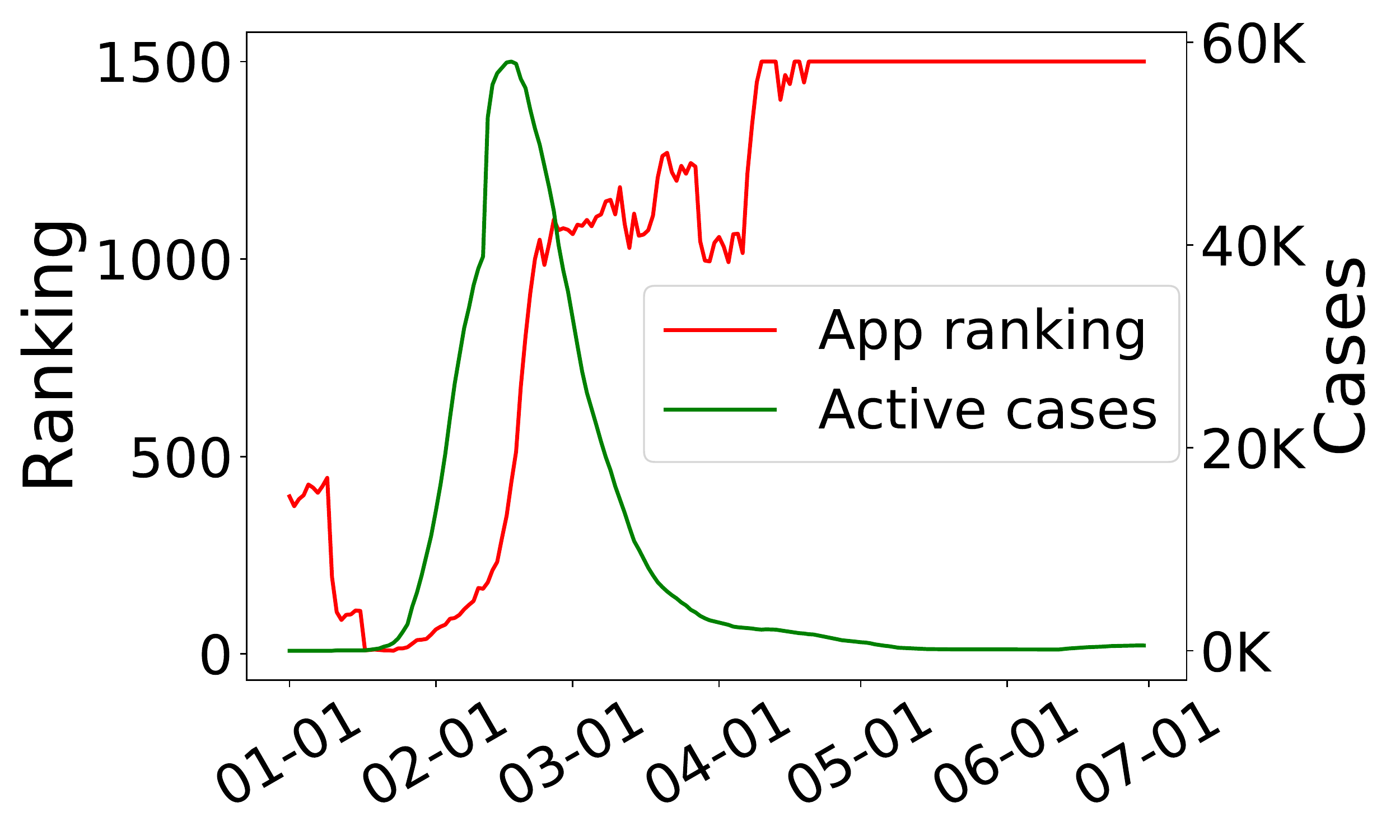}}
    \vspace{-0.15in}
    \caption{Two examples for SNC with the decline stage in China.}
    \label{fig:c6-example}
\end{figure}

As expected, this category contains apps whose rankings have a strong negative correlation with the active cases only in the decline stage, i.e., as the pandemic improves, the ranking declines. For this case, some apps may seem less important as people gradually return to their normal lives. For example, the app with the highest correlation coefficient (-0.852) is \textit{Fast View}, an informational app that aggregates and provides users different forms of content including graphics, videos, etc., as well as current affairs content and the latest real-time information.
Furthermore, there is an online gaming app called \textit{Doudizhu} (-0.762), which is a very popular card game in China. This rose to the top of the rankings at the beginning of the outbreak, but quickly declined for a long time as shown in Figure~\ref{fig:doudizhu}.
These apps seem to become popular for a short period of the outbreak as a means for people to keep abreast of events or to pass the time. However, they cannot last very long, especially when people return to their normal life routines.

\subsubsection{SPC with growth stage, SNC with decline stage \& SNC with growth stage, SPC with decline stage.}
\begin{figure}[htb]
    \centering
    \subfigure[Cube Battle]{
    \label{fig:cubebattle}
    \includegraphics[width=0.4\textwidth]{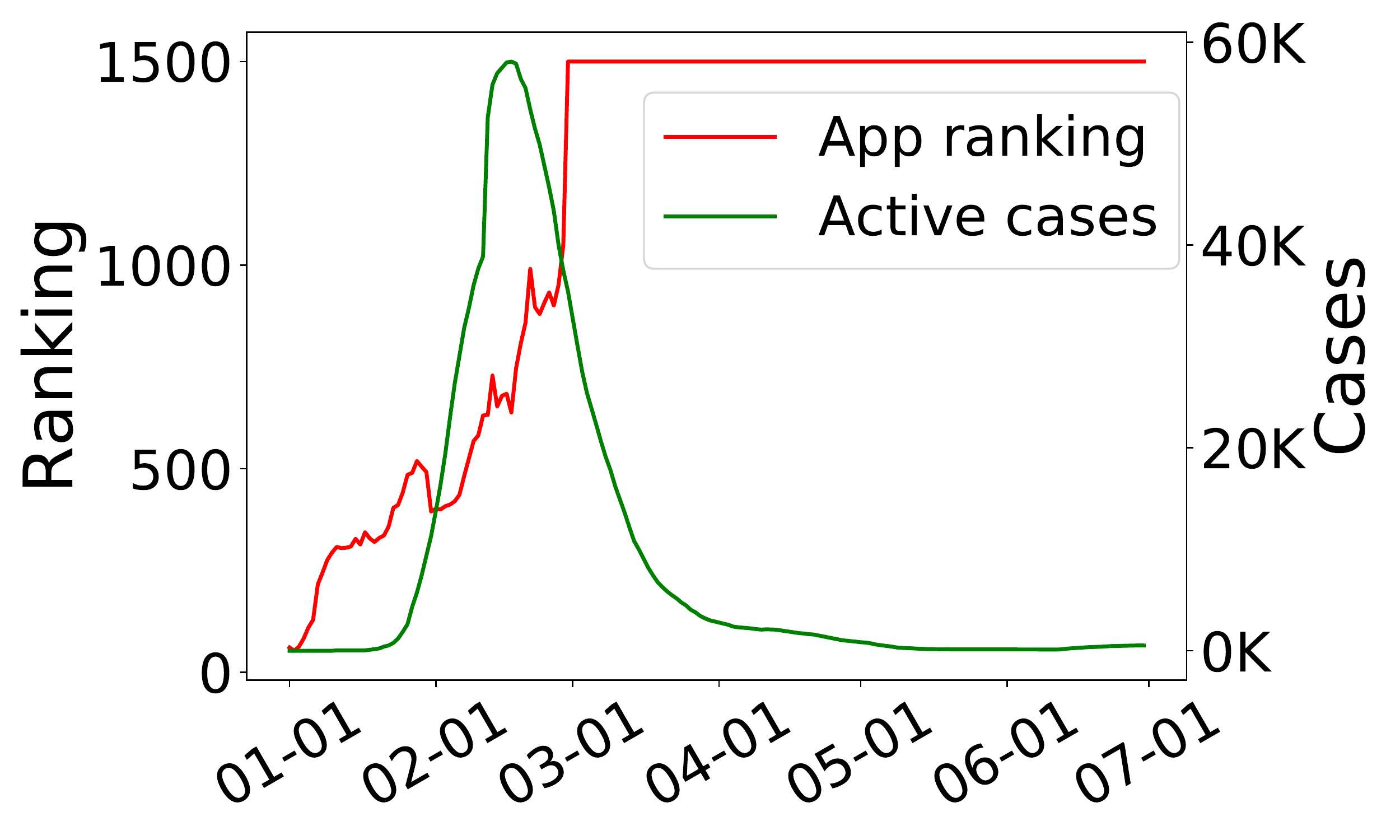}}
    \quad
    \subfigure[SoYong]{
    \label{fig:soyoung}
    \includegraphics[width=0.4\textwidth]{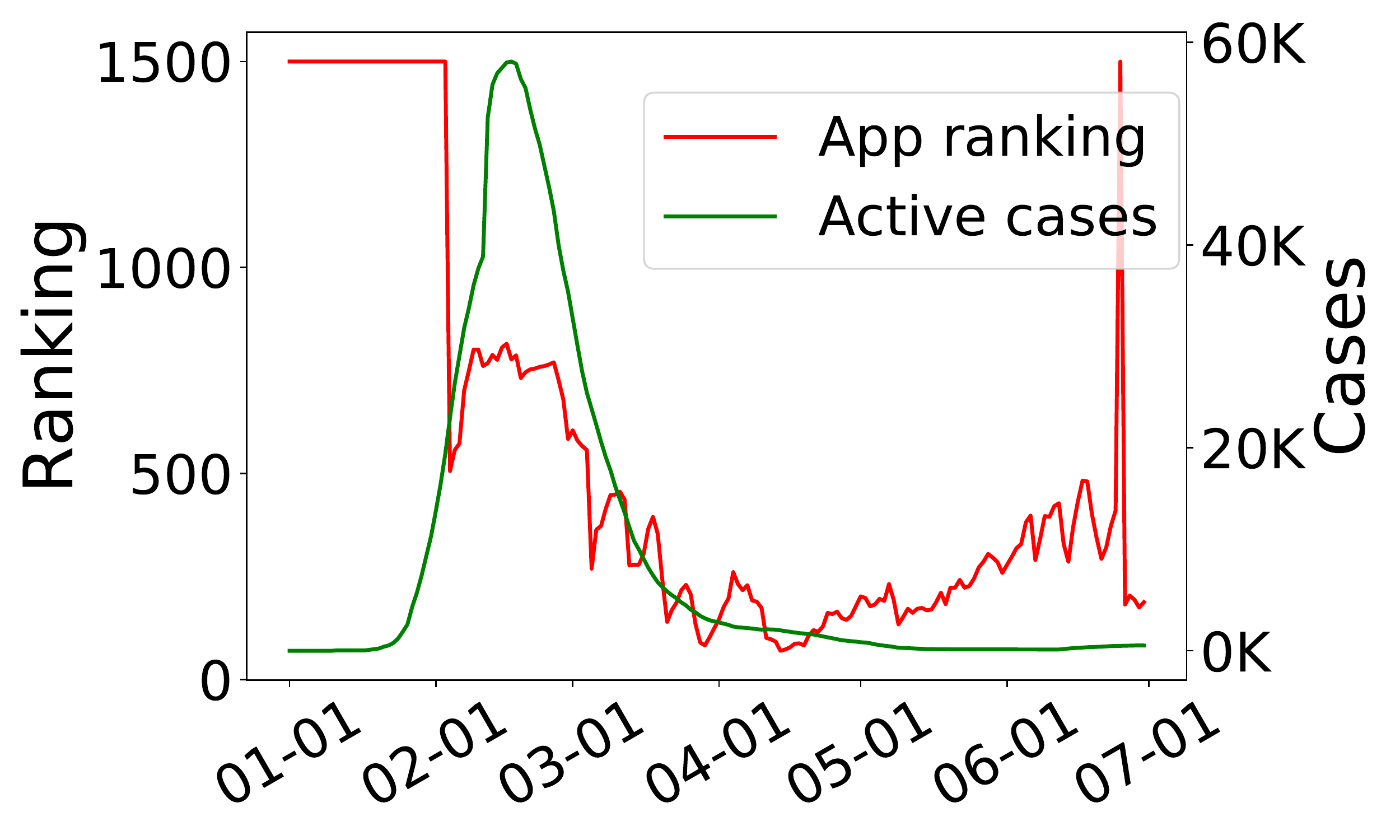}}
    \vspace{-0.15in}
    \caption{Two examples for the two-stages correlations respectively.}
    \label{fig:c7-example}
\end{figure}

In addition to the above apps, there are also a few apps that have strong correlations with both stages but in different directions. There are 7 apps that show strongly positive correlations during the growth stage of the pandemic while exhibiting negative correlations with the decline stage. Their rankings can be roughly treated as constantly declining over the outbreak. For example, there is a puzzle elimination game called \textit{Cube Battle} that has fallen steadily out of the top 1,500 since the outbreak began, as shown in Figure~\ref{fig:cubebattle}. It has a high correlation coefficient of 0.803 within the growth stage and -0.878 within the decline stage. In contrast, the 2 apps that are strongly negatively correlated with the growth stage and positively related to the decline stage exhibit the opposite trend, i.e., the ranking is almost continuously rising for most of the time. For example, there is a beauty and cosmetic app called \textit{SoYong} that has a generally fluctuating upward trend in the ranking from February to May, as displayed in Figure~\ref{fig:soyoung}. It has the correlation coefficient of -0.86 within the growth stage and 0.717 within the decline stage. Note that both types are few in number, indicating that the two evolving patterns may be very uncommon.

\subsubsection{SPC and SNC in the US.}
\begin{figure}[htb]
    \centering
    \subfigure[Trivia Crack]{
    \label{fig:triviacrack}
    \includegraphics[width=0.4\textwidth]{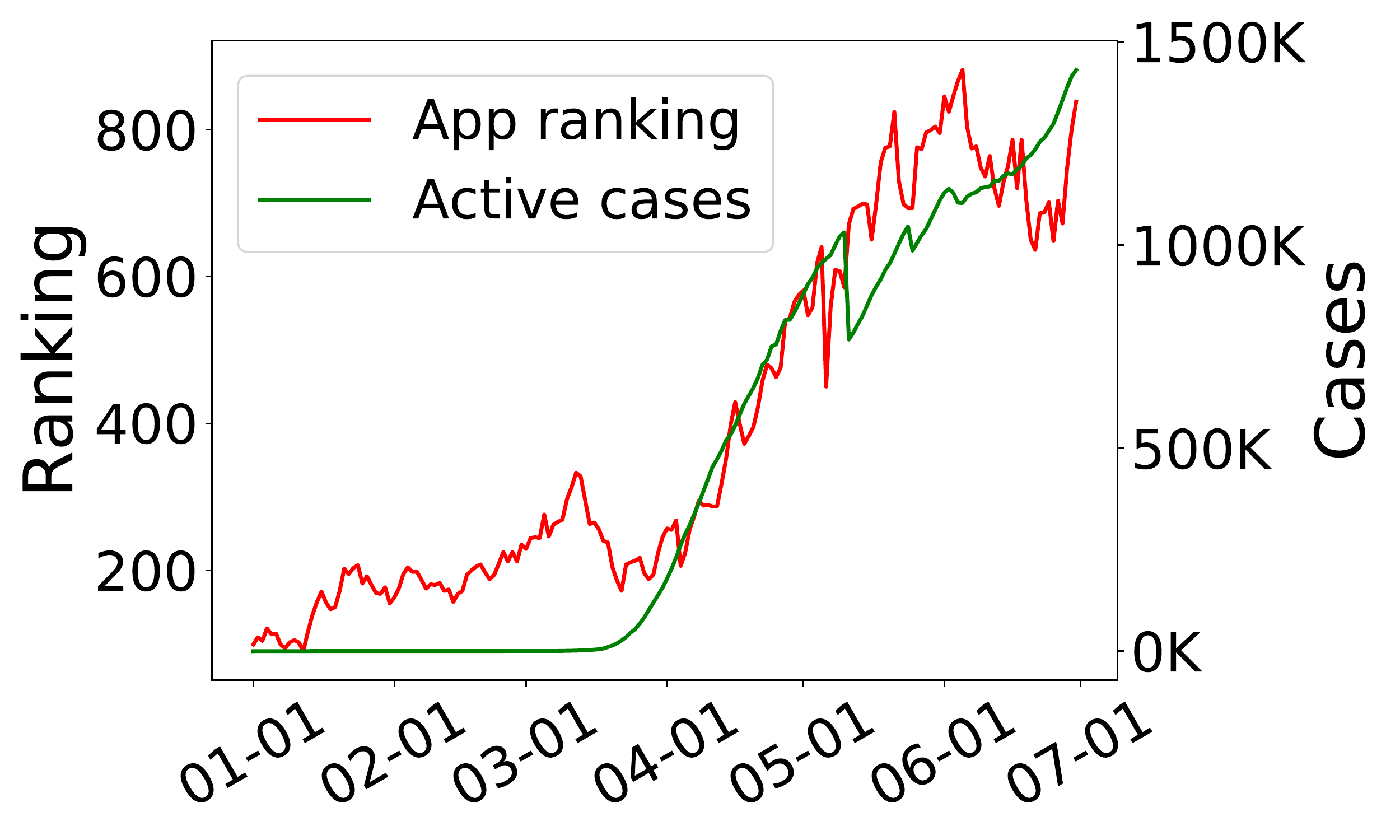}}
    \quad
    \subfigure[Videoleap Video Editor \& Maker]{
    \label{fig:videoleap}
    \includegraphics[width=0.4\textwidth]{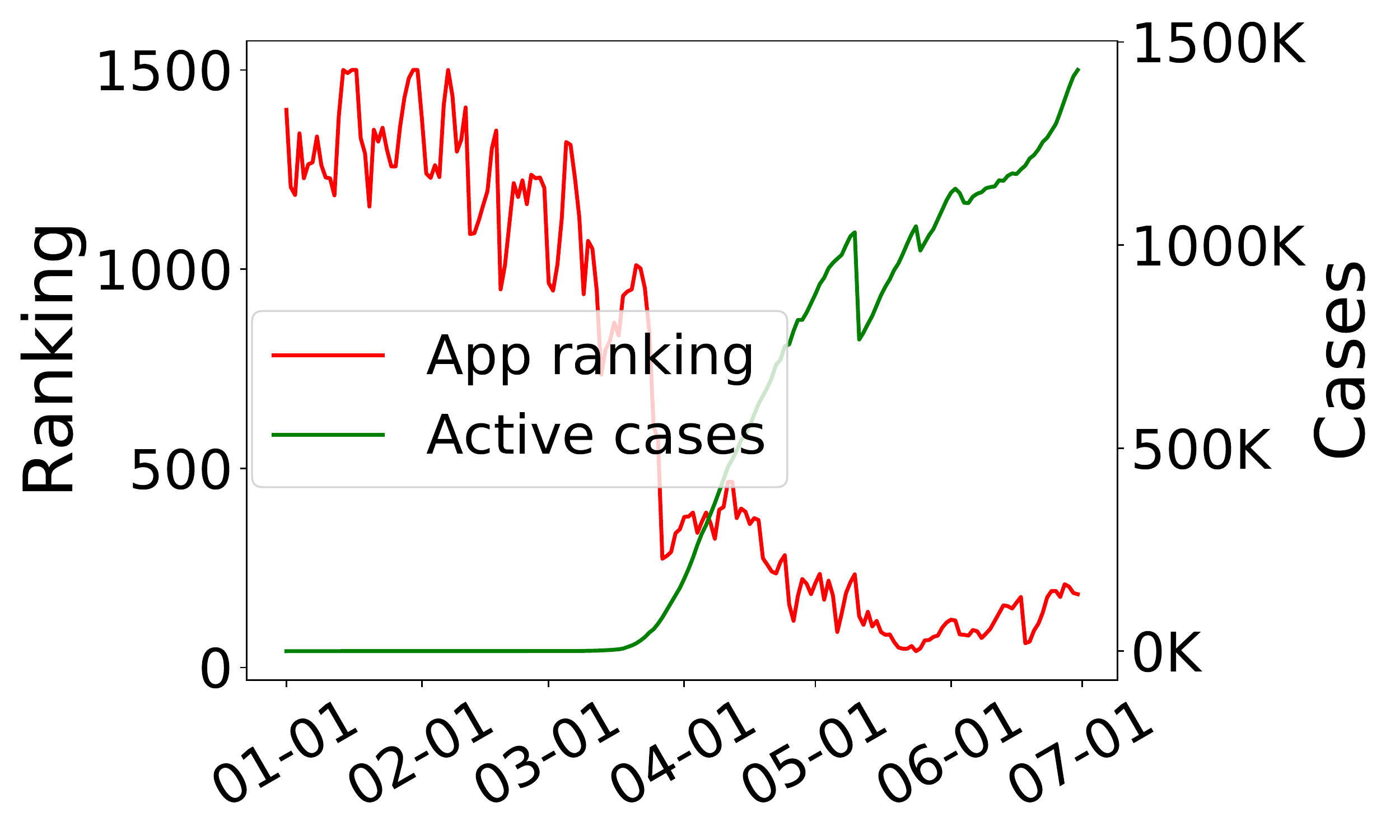}}
    \vspace{-0.15in}
    \caption{Two examples for the US.}
    \label{fig:us-example}
\end{figure}

Due to the single trend in the number of the active confirmed cases in the US, i.e., consistently increasing since mid-March, we construct two categories: SPC and SNC with the overall, with 72 and 116 apps, respectively. Since the pandemic situation in the US has been increasingly serious, the apps with an SPC to the pandemic case rate should show a largely declining ranking. For example, the app with the highest correlation coefficient (0.965) is a gaming app called \textit{Trivia Crack}, which has its ranking change in parallel with the number of active cases, as shown in Figure~\ref{fig:triviacrack}. In contrast, the apps with an SNC with the pandemic display a generally rising trend. An example is displayed in Figure~\ref{fig:videoleap}, which is an app called \textit{Videoleap Video Editor \& Maker} (-0.88), which is an app for producing videos. Its ranking has kept advancing during the six months.  

%\subsection{Summary}
%In terms of the classification observations, apps' rankings exhibits different evolution processes with a great relation to the pandemic.

\subsection{Summary}

In summary, the pandemic has diverse patterns in terms of its impact on app ranking. In China, 85 out of 586 apps are strongly correlated with the half-year pandemic, either positively or negatively. Moreover, 111 apps have strong correlations (positive or negative) with only one stage of the pandemic (growth or decline). 9 apps even show strong correlations with the two stages in different directions. These correlation categories reflect different characteristics of the COVID-19 effect on app rankings. Most of the apps in the same category share some similarities in terms of their adaptability to the pandemic situation.
While for the US, since the vast majority of apps are game apps, this somewhat limits our study of the adaptation of app functions and features to the pandemic.

\section{Characterizing Side Effects}
\label{sec:sideeffect}

\begin{figure}[htbp]
    \centering
    \subfigure[ViLin]{
    \includegraphics[width=0.4\textwidth]{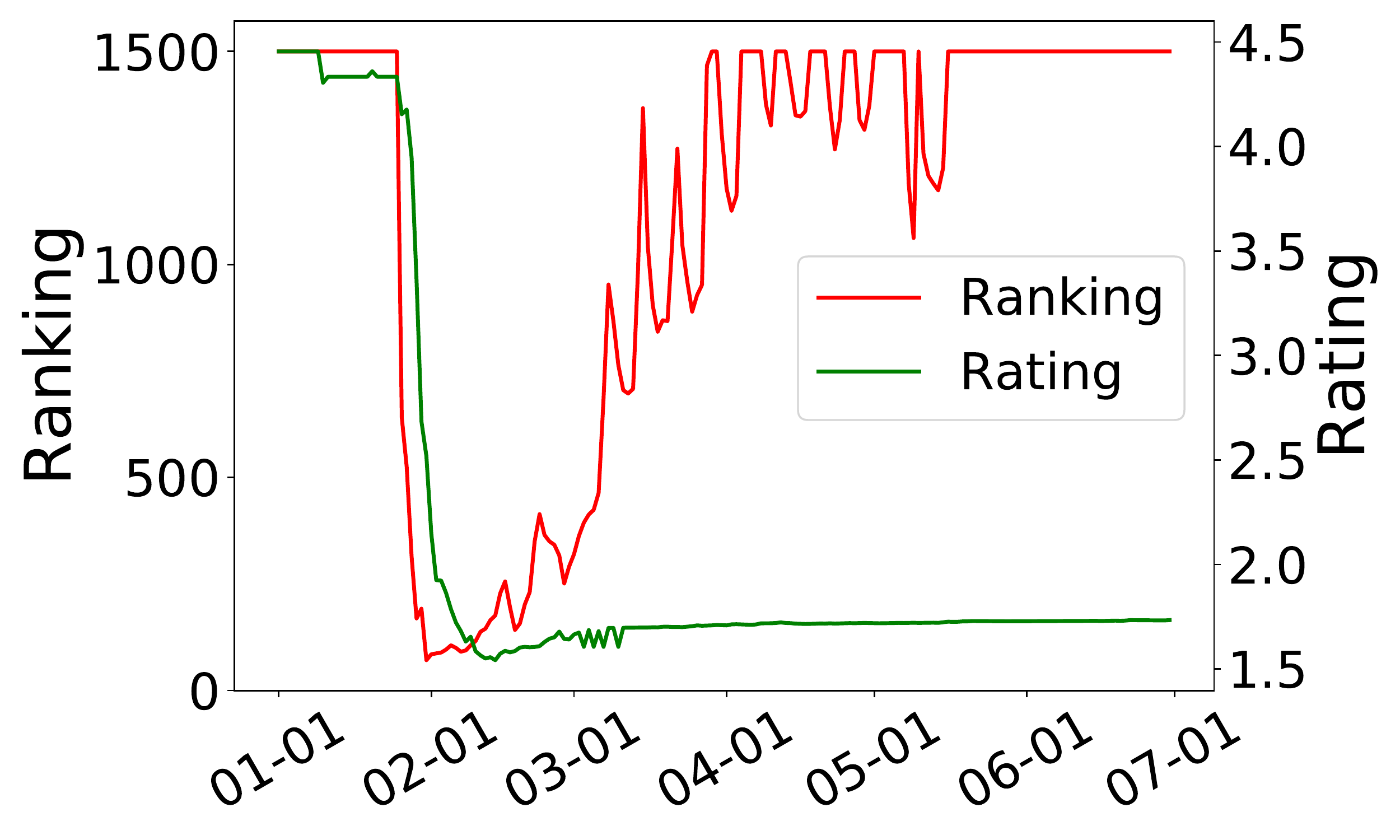}}
    \quad
    \subfigure[Tencent Meeting]{
    \includegraphics[width=0.4\textwidth]{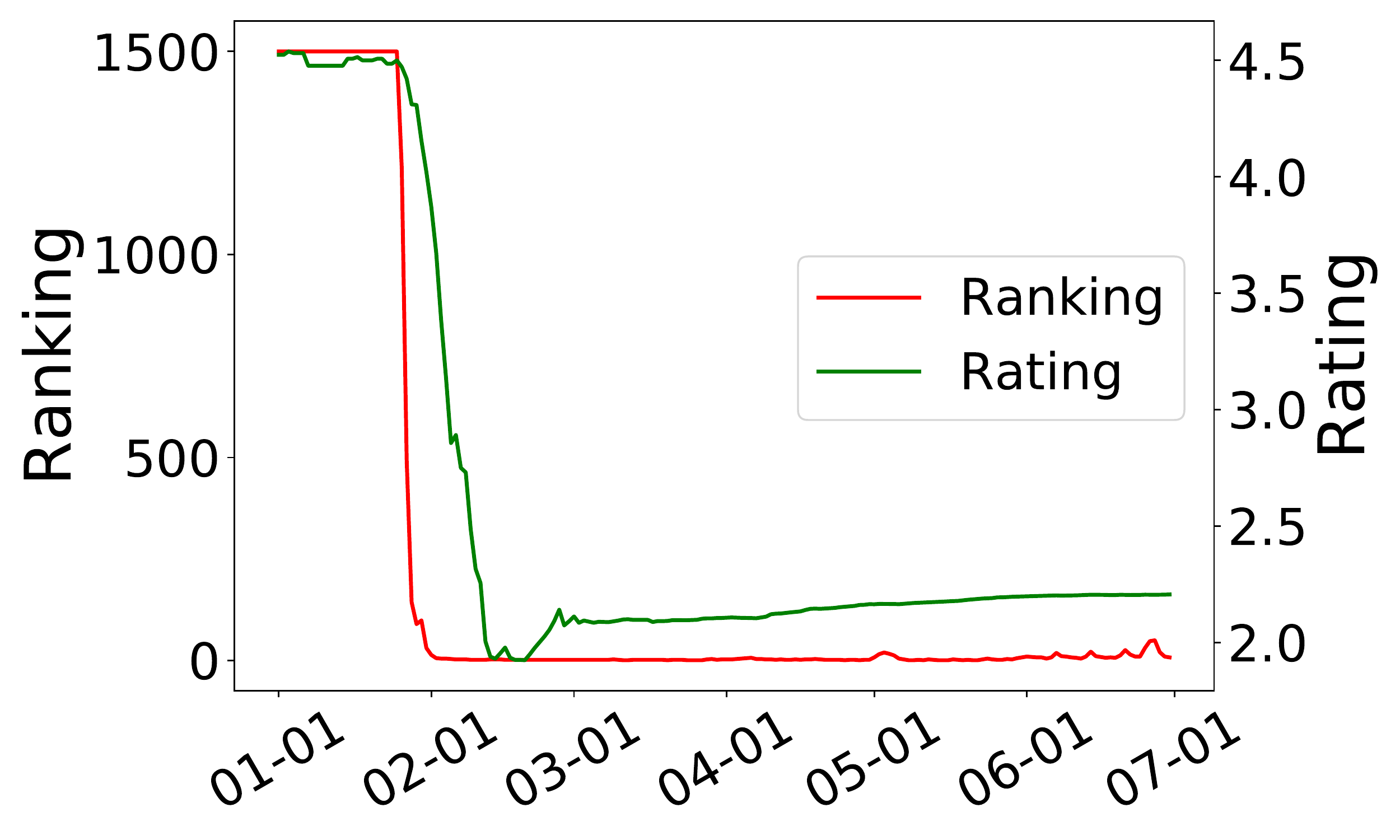}}
    \vspace{-0.15in}
    \caption{Two app examples where the rating goes down as the ranking goes up.}
    \label{fig:rank-rate-example}
\end{figure}

We conclude by briefly inspecting side effects that may be caused by these rapid fluctuations in app popularity. Particularly, an interesting phenomenon we observe is that the ratings of some apps are likely to decrease in varying degrees as their rankings go up.
For example, Figure~\ref{fig:rank-rate-example} shows the trends in ranking and rating of the two app examples over this period. 
While the ranking and popularity of the app is rising rapidly, its rating is experiencing a decline.
This prompts us to investigate whether the rise in app ranking and popularity may have some side effects on app maintenance behavior.

\subsection{Ratings Analysis}
We briefly focus on those apps that gained popularity for a short or long time as COVID-19 hit, exploring the relationship between their ratings and rankings.
Considering that there is a wide diversity in the evolution of many app rankings throughout the time span, we shorten the time frame to the period when COVID-19 was first spreading, and target only those apps whose rankings rose rapidly during this period.

Specifically, for \emph{China}, we choose the period when the ranking exhibits strong negative correlation with the number of active confirmed cases during the worsening of the COVID-19, i.e., from January to mid-February. 
For the \emph{US}, we pay attention to the two months just after the outbreak of COVID-19, i.e., from March to May.

We then calculate the correlation coefficients between the ratings and rankings of the focused apps. 
The results are presented in Table~\ref{tab:rank-rate}. We see that there were 95 apps in China and 88 apps in the US that improved their rankings significantly at the beginning of the COVID-19 outbreak. Of these, 63 apps in China and 24 apps in the US have a correlation coefficient \emph{r}>0 between rating and ranking.
This means that their ratings decreased as their ranking rises, i.e., introducing side effects. 
Curiously, the results show that this particle side effect is more prevalent in China than the US. 

\begin{table}[htbp]
    \centering
    \caption{Distribution of correlation coefficient between rankings and ratings for focused apps.}
    \begin{tabular}{c|c|c|c}
    \toprule
      Country  & \# Apps & \# r<0 (p<0.05) & \# r>0 (p<0.05)\\
      \hline
      China & 95 & 29 & 63 \\
      US & 88 & 60 & 24\\
      \bottomrule
    \end{tabular}
    \label{tab:rank-rate}
\end{table}

\subsection{Reviews Analysis}
We have shown that some apps that gains popularity while experiencing a decline in their ratings. 
We next dive into the reasons behind this phenomenon and investigate the challenges for app developers.  
Previous studies~\cite{carreno2013analysis,pagano2013user} showed that reviews represent a rich source of information for app developers, such as user requirements, bug reports, feature requests, and documentation of user experiences. Thus, we look at user reviews to analyze the possible reasons for the drop in app ratings. 
It is important to note that while the ratings of all of these apps have dropped, the magnitude of the drop varies greatly, with many showing only slight drops.
For the sake of representativeness, we target apps where the correlation coefficient between app rankings and ratings is greater than 0.6 and the absolute value of the range of rating decline is greater than 0.1. This leaves 37 apps for review analysis.

We use AppBot to extract the app reviews. This tool provides a large number of data-mining and sentiment analysis features, and is used in many studies, e.g.,~\cite{garousi2020mining}.
For each app, we use AppBot for automated text mining and sentiment analysis of reviews.
To uncover new traits, we use reviews from the second half of 2019 as the benchmark for comparision.  

Overall, we find a significant increase in the number of reviews for these apps in the first half of 2020 compared to the second half of 2019. This is intuitive considering their increased demand. A total of 1,355,460 reviews occur in the first half of 2020, which is 4.3 times more than the 313,209 reviews in the previous six months.
More importantly, sentiment analysis reveals that the number and percentage of negative reviews among them has also increased considerably, with a total of 716,129 negative reviews in the first half of 2020 (53\%) compared to 72,345 in the second half of 2019 (23\%), a visible number about 10 times higher. 
This is also the case for individual apps. For example, \textit{Tencent Meeting} had only 33 user reviews during the second half of 2019, of which only 5 (15\%) were negative, while the number of reviews during the first half of 2020 was astonishingly high at 10,507, of which 4,583 (44\%) were negative.
This may mean that such apps have struggled to fulfil user expectations as their user base has grown. 

Further, we focus on the negative reviews and seek to get a sense of the topics of user complaints. 
We thereofre rely on AppBot's topic grouping feature and identify six core topics raised within these reviews:

(1) \emph{Bugs.} There are lots of negative reviews describing problems with the app which should be corrected, such as a crash, an erroneous behavior, or a performance issue. This is likely a product of more users experimenting with the features of the apps.

(2) \emph{Device.} There are many negative comments regarding the device, including device incompatibility, interface mismatch, lack of matching function and device features, unfriendly to ipad devices, triggering device heat and lag, etc. Again, such issues become more prevalent as your user base diversifies. 

(3) \emph{Connectivity.} This mainly refers to network connection problems, such as difficult WiFi connections, slow loading, long waiting times, etc.

(4) \emph{Design \& UX.} There are some negative reviews about the user interface, specifically that the user interface is not attractive, and the UI design and interaction are not friendly enough, making the user experience poor.

(5) \emph{Video/Audio.} For some apps, there are some problems with opening video or voice, such as delay, not smooth, etc. Usually such complaints arise in apps featuring multi-person online real-time sessions, such as online meetings and online lecture apps.

(6) \emph{Privacy.} There are also some concerns about privacy and security of some apps, such as invasion of privacy, sharing information with third-parties, stealing edits and tracking location, etc.

\begin{figure}
    \centering
    \includegraphics[width=0.9\textwidth]{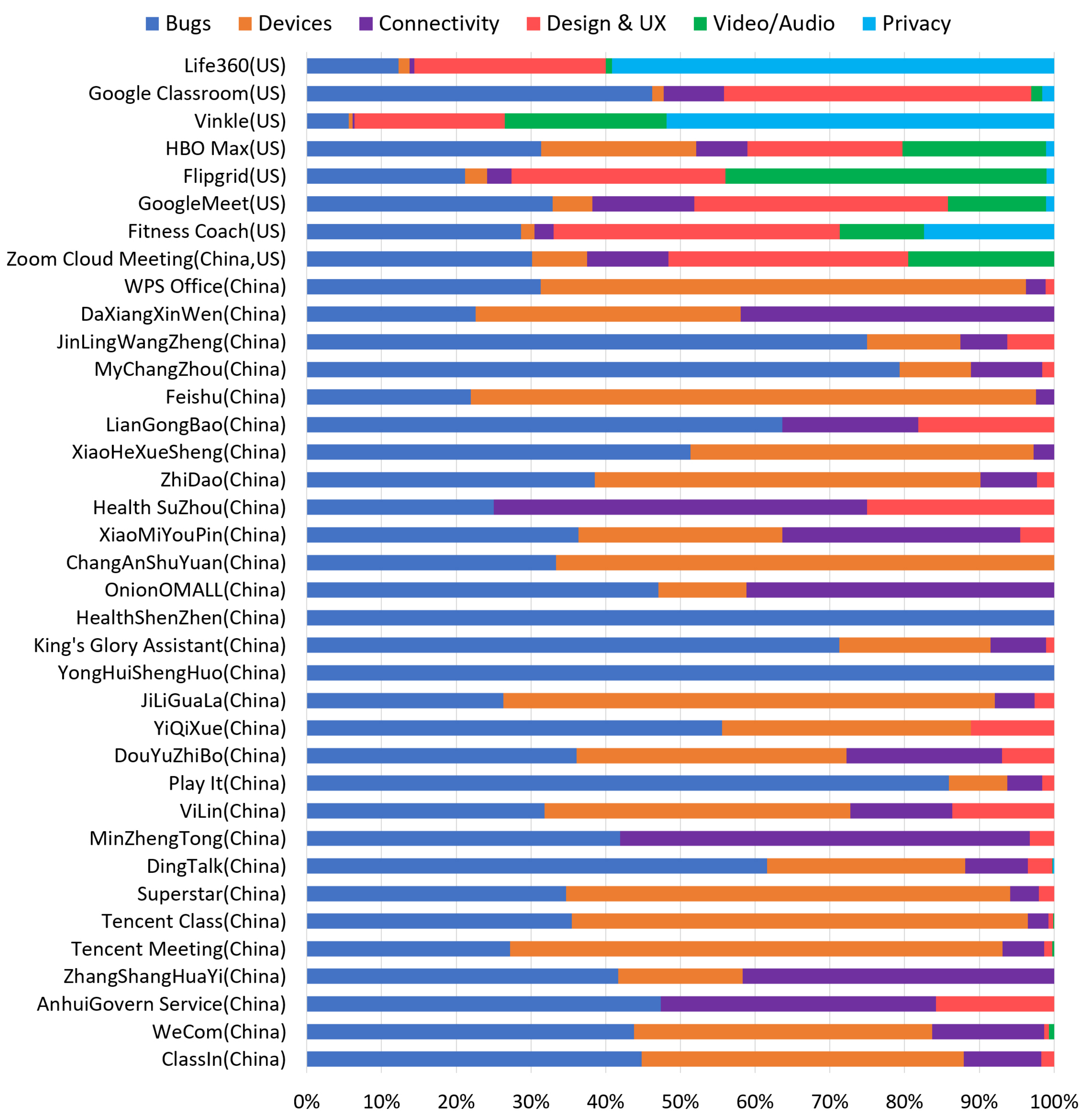}
    \caption{Distribution of the six negative review topics in each app.}
    \label{fig:app-review}
\end{figure}

We further measure the distribution of different kinds of topics for each app, as shown in Figure~\ref{fig:app-review}.
As can be seen, the distribution of user concerns varies greatly across apps, where bugs and devices issues are prevalent.
We also notice that some of the issues are related to the increase in the number of users. For example, \textit{Play It}, a multiplayer interactive trivia game app in China, suffered several crashes due to a surge in the number of users during the initial outbreak of COVID-19 in China as people were quarantined at home.
Generally speaking, as more people use the app, the number of complaints grows as well. Some issues that used to be obscure seem to be magnified, such as the aesthetics of the user interface.
This sudden popularity presents a great challenge to the app in many aspects including functionality, compatibility, stability, fault tolerance and UI design, etc.
In a way, it serves as a wake-up call for developers to learn from the experience in dealing with unexpected situations in the future. Meanwhile, it offers unprecedented opportunities for app developers to resolve the hidden problems.

\section{Discussion}

\subsection{Implications}
The most significant result in our paper is that COVID-19 has had a significant impact on the app rankings in app store. This impact is reflected in both the popularity of several categories and the rankings of individual apps. Moreover, the evolution of app rankings are diverse and have a lot to do with whether the features and functionalities of the apps are compatible with the current environment. Besides, we notice that some apps that rapidly rise in popularity may also experience the side effect of rating drops. This reflects the challenges that rapid changes in popularity can cause app developers. We also note that findings are different between China and the United States.

The relevant stakeholders should take note of these findings, as they play different roles at different levels of the app market. Our findings have further implications for understanding the behavior of the mobile app ecosystem during public health crises, and helps developers to make better decisions on app developmennt and management. As our first finding indicates, the popularity of apps may experience disturbances resulting from public events. Such disturbances demonstrate the importance of having a contingency plan. When some sudden public events such as the COVID-19 pandemic happens, a contingency plan could help app developers, operators, as well as managers better cope with them. For example, one may develop a quick server deployment plan when expecting high demand. App developers may diversify their portfolios to avoid potential loss caused by popularity decreases. The second finding suggests that the impact may be complicated, depending on how the event is going on. Thus, app developers may need to develop the ability to develop the short-term and long-term insights, and plan accordingly. For example, for an app that gains popularity at the early phase of the pandemic but gradually return to normal later, app developers need to decide if it is necessary to invest resources for the short-term spikes. We also notice that increases of app popularity often coincide with decreases in ratings, and growing number of reviews containing users' insights. App developers thus need to be vigilant to users' feedback, and take this opportunities to improve the app quality. Moreover, our results and findings are not restrict to a specific event; it could be helpful in other public emergencies at the global scale.

.

\subsection{Limitations}
We recognize that our study carries several limitations and threats to validity.
First, since the main impact of the pandemic was reflected on only a small number of apps, we selected the top 100 apps each day, yet there was no accepted standard for selecting this number. 
Second, in determining the strength of the correlation, we set a threshold value of 0.6. However, there is no standard definition for the selection of this threshold. 
Third, only a small number of apps have been used for our review analysis. The major reason is that many apps have insignificant side effects. Hence, we are concerned that reviews of such apps are not representative and reliable enough.

\section{Conclusion}
This paper has presented the first longitudinal empirical study of the evolution of app rankings during COVID-19. Our analysis covers 586 apps in China and 590 apps in the US with records of being ranked in the top 100 from January 1 to June 30, 2020, across 22 categories. We performed analysis from the perspectives of category popularity. To further understand the evolution patterns of app rankings, we proposed a correlation-based method to classify them and then perform a per category analysis. Besides, we characterized the side effect of declining ratings accompanied by the rising popularity of some apps. Our observations reveal mobile users' reactions in the mobile ecosystem in the face of unexpected public crises, and provide insights for app developers to make better decisions on developing apps.

\balance

\bibliographystyle{plain}
\bibliography{cite}

\end{document}